\documentclass{article}
\usepackage{amsfonts}
\usepackage{amsmath}
\usepackage{amsthm}
\usepackage{multirow}
\usepackage{rotating}
\usepackage{booktabs}
\usepackage{hyperref}

\begin{document}

\def\mH{\mathbb{H}}
\def\mC{\mathbb{C}}
\def\mP{\mathbb{P}}
\def\mR{\mathbb{R}}
\def\mQ{\mathbb{Q}}
\def\mZ{\mathbb{Z}}
\def\mN{\mathbb{N}}

\newtheorem{theorem}{Theorem}
\newtheorem{lemma}{Lemma}
\newtheorem{proposition}{Proposition}
\newtheorem{assumption}{Assumption}{\bf}{\it}
\newtheorem{conjecture}{Conjecture}

\def\sqr#1#2{{\vcenter{\vbox{\hrule height.#2pt
        \hbox{\vrule width.#2pt height#1pt \kern#1pt
          \vrule width.#2pt}\hrule height.#2pt}}}}
\renewcommand\square{
  \mathop{\mathchoice{\sqr{20}{25}}{\sqr{12}{15}}{\sqr{8}{10}}{\sqr{4}{5}}}}
\def\orbox#1#2{{\scriptstyle{#1}}\square_{#2}\limits} 

\def\cADE{$\mathcal{A}$-$\mathcal{D}$-$\mathcal{E}\,$}

\title{On the complete classification of unitary $N=2$ minimal superconformal field theories}
\author{Oliver Gray\\
  {\it Department of Mathematics}\\
  {\it University of Bristol, Bristol, BS8 1TH, UK}\\
\texttt{oliver.gray@bristol.ac.uk}}
\maketitle

\begin{abstract}
  Aiming at a complete classification of unitary $N=2$ minimal models (where the assumption of space-time supersymmetry has been dropped), it is shown that each modular invariant candidate partition function of such a theory is indeed the partition function of a fully-fledged unitary $N=2$ minimal model, subject to the assumptions that orbifolding is a `physical' process and that the space-time supersymmetric \cADE models are physical.  A family of models constructed via orbifoldings of either the diagonal model or of the space-time supersymmetric exceptional models then demonstrates that there exists a unitary $N=2$ minimal model for every one of the allowed partition functions in the list obtained from Gannon's work~\cite{Gannon1996}.

  Kreuzer and Schellekens' conjecture~\cite{Kreuzer1994} that all simple current invariants can be obtained as orbifolds of the diagonal model, even when the extra assumption of higher-genus modular invariance is dropped, is confirmed in the case of the unitary $N=2$ minimal models by simple counting arguments.
\end{abstract}

\section{Introduction}
Conformal field theories (CFTs)~\cite{Belavin1984,Ginsparg1988,DiFrancesco1997,Gaberdiel1998,Gaberdiel1999} have been a well-studied area of research since they first became a hot topic following the publication of the seminal paper of Belavin, Polyakov and Zamolodchikov in 1984~\cite{Belavin1984}.  In their paper, the authors laid down the formalism of conformal field theories by combining the representation theory of the Virasoro algebra with the concept of local operators, and discovered the minimal models.  The term \emph{minimal} indicates that the Hilbert space of the CFT decomposes into only finitely many irreducible representations of (two commuting copies of) the Virasoro algebra.  The existence of null-vectors in the Hilbert spaces of minimal models permit ODEs for the correlation functions to be derived, which in turn allow the minimal models to be completely solved.  Miraculously, the minimal models turned out to describe phenomena in statistical mechanics~\cite{Cardy2008}; most notable is their description of 2nd or higher order phase transitions, e.g. the Ising model~\cite{Onsager1943,Belavin1984} and the tri-critical Ising model~\cite{Friedan1984b}.

Once the inequivalent irreducible unitary representations of the Virasoro algebra with central charge $0\leq\mathbf{\bar{c}}<1$ were known, the next problem was to piece them together in a modular invariant way (see section~\ref{mi}).  All modular invariant combinations were found~\cite{Cappelli1987} to fall into the well-known \cADE meta pattern (see e.g.~\cite{Zuber2000}).

The classification of other classes of conformal field theories has been the aim of much work, and is an ongoing project.  Most promising is the study of \emph{rational} theories, whose Hilbert spaces may contain infinitely many irreducible representations of the Virasoro algebras, but which can be organised into a finite sum of representations of some larger so-called \emph{$W$-algebra}.  An important source of rational theories are the WZW models~\cite{Wess1971,Witten1983}: families of theories, which can be constructed for any semi-simple finite-dimensional Lie algebra $\mathfrak{g}$.  Many of the families of WZW models have been at least partially classified~\cite{Gannon1994b,Gannon1994c}, the most famous being the complete classification of the $\mathfrak{g}=\mathfrak{su}(2)$ case~\cite{Cappelli1987}, which again correspond to the \cADE series.

Another source of rational CFT is inspired by string theory~\cite{Becker2007,Green1987,Green1987a,Polchinski2005}, the most promising candidate for a description of the fundamental forces of the universe.  String theorists have developed the notion of supersymmetry, the idea that there is a symmetry between bosonic and fermionic matter in our universe.  In mathematical terms, the Virasoro algebra is enlarged by adding $N$ supersymmetry operators (and their super partners).  One can then consider superconformal field theories (SCFTs), theories that fall into representations of this enlarged algebra.  The minimal unitary $N=2$ superconformal field theories~\cite{BFK,DiVecchia1986,DiVecchia1986b,Yu1986,Matsuo1987,Kiritsis1987,Ravanini1987,Qiu1987,Qiu1987a} (or unitary $N=2$ minimal models), for example, provide building blocks for Gepner models (see e.g.~\cite{Greene1996}).

Contrary to popular belief, to date the unitary $N=2$ minimal models have not been completely classified.  It is commonly stated that they also fall into the \cADE meta-pattern, due to the work of~\cite{Martinec1989,Vafa1989,Cecotti1993}, in which those unitary $N=2$ minimal models that enjoy space-time supersymmetry are demonstrated to be in one-to-one correspondence with the \cADE simple singularities.  But when one quite reasonably drops the condition of space-time supersymmetry, one finds a much larger possible set of solutions.

The condition of space-time supersymmetry means that there should be a fundamental symmetry between space-time bosons and fermions; in a SCFT, the symmetry implies that all information about the space-time anti-periodic fields (the R sector) is encoded by the space-time periodic fields (the NS sector) and vice-versa.  This relation is encoded by the \emph{spectral flow} (see e.g.~\cite{Greene1996} and section~\ref{spacetime}), which provides an explicit map from one sector to the other in supersymmetric theories.

Gannon~\cite{Gannon1996} classified the possible partition functions of the unitary $N=2$ minimal models, showing that in fact there is a much larger playground than previously suspected: there are finitely many partition functions at each level $k$, but the number is unbounded as $k$ increases, in contrast with the $N=0$ case.  There are also many more ``exceptional'' cases: $10,18$ and $8$ corresponding to what are somewhat misleadingly termed the $\mathcal{E}_6,\mathcal{E}_7$ and $\mathcal{E}_8$ models, respectively.

Two natural questions then arise: do all of these partition functions belong to genuine SCFTs, or are some just mathematical curiosities?  And could there be more than one minimal model associated to each partition function?  In this paper we attempt to answer the first of these questions.  Perhaps surprisingly, it can be resolved using only orbifold-related arguments, in the following way: orbifoldings~\cite{Dixon1985,Dixon1986} from every possible partition function to the partition function of one of a small list of well-known and fully understood models are explicitly calculated.  Coupled with the widely-held beliefs that (i) the space-time supersymmetric $\mathcal{A},\mathcal{D}$ and $\mathcal{E}$ models are fully-fledged, physical SCFTs and (ii) orbifolds of a physical theory are again physical, this demonstrates that each partition function is indeed that of a physical SCFT.  This is an important step towards the full classification of the unitary $N=2$ minimal models.

We note that Kreuzer and Schellekens~\cite{Kreuzer1994} have proven a related result.  They construct simple current modular invariant partition functions via orbifoldings of the diagonal model and use the further assumption of higher-genus modular invariance to show that all simple current modular invariant partition functions can be obtained this way.  They hypothesise that this extra assumption is unnecessary, which we are able to confirm for the case of unitary $N=2$ minimal models by simple counting arguments.

Section 2 is a review of Gannon's program of classifying the possible partition functions of the $N=2$ unitary minimal SCFTs, and the statement of his result (which did not appear explicitly in~\cite{Gannon1996}), with a few minor errors corrected.  We give two examples to illustrate the simplest cases in the classification.

In section 3 we show that the \cADE classification of Cecotti and Vafa~\cite{Cecotti1993} is visible, even at the level of partition functions.  We give a conjecture for the fusion rules of the $N=2$ minimal models and give a much simpler proof than the one found in~\cite{Wakimoto1998} that they follow from Verlinde's formula.  Then we prove results showing that Gannon's partition functions are compatible with the conjectured fusion rules and with the locality of the associated theories.

Section 4 contains a brief review of orbifold techniques, and the statement and proof of the main theorem: every possible partition function in section~\ref{minimalmodels} belongs to a fully-fledged SCFT (subject to the assumptions (i) and (ii) above).  The proof is an explicit construction of orbifoldings from any given partition function to one of a handful of fixed theories that are believed to be fully-fledged SCFTs.

Section 5 investigates the simple-current modular invariants and confirms a hypothesis of Kreuzer and Schellekens for the special case of the unitary $N=2$ minimal models; namely, that every simple current invariant should be obtainable via an orbifold of the diagonal model.

Section 6 contains conclusions and further directions to be investigated.

\section{Gannon's Classification of Partition Functions}
\subsection{Preliminaries}
We will denote by $\mH$ the underlying pre-Hilbert space of an $N=2$ SCFT $\mathcal{C}$.  $\mH$ is a representation of two commuting copies of the $N=2$ super Virasoro algebra (SVA)~\cite{Ademollo1975}, whose `modes' are $1,L_n,J_n,G^{\pm}_r$ with $n\in\mZ$ and $r\in\mZ+\frac{1}{2}$ in the Neveu-Schwarz (NS) sector and $r\in\mZ$ in the Ramond (R) sector.  The $L_n$ modes along with the central element $1$ form a Virasoro algebra with central charge $\mathbf{\bar{c}}\in\mC$, the $J_n$ are the modes of a $U(1)$ current\footnote{Our normalisation of the $U(1)$ current agrees with that of e.g.~\cite{Qiu1987a}.  As a consequence, $[J_0,G^{\pm}_r]=\pm\frac{1}{2}G^{\pm}_r$, and so the supersymmetry modes $G^{\pm}_r$ carry half integer charge.}, and the $G^{\pm}_r$ are modes of two fermionic super-partners.  Together these elements span the left-hand copy of the SVA.  The right-hand copy of the SVA is spanned by the elements $\{\overline{1},\overline{L}_n,\overline{J}_n,\overline{G}^{\pm}_r\}$ with the same commutator relations.

Unitary irreducible inequivalent representations of the SVA can be realised as lowest weight representations (LWRs)\footnote{Lowest weight representations are frequently referred to as highest weight representations, a slightly perverse accident of history given that the `highest weight vector' actually has the lowest weight of all states in the representation.} which are characterised by a lowest weight vector $v$ with lowest weight $h$ and charge $Q$:
\begin{align*}
  &L_0v=hv,\quad J_0v=Qv,\\
  &L_nv=J_nv=G^{\pm}_rv=0 \quad\forall n>0,\;r>\frac{1}{2}.
\end{align*}
Through calculation of the vanishing curves of the Kac determinant, Boucher, Friedan and Kent~\cite{BFK} classified these irreducible unitary representations.  They exist only when
\begin{align}
  \mathbf{\bar{c}}=\frac{3k}{\overline{k}},\qquad k\in\mN_0=\{0,1,2,\ldots\}\label{c}
\end{align}
where throughout the paper we will write
\begin{align}\label{kbar}
  \overline{k}&=k+2.
\end{align}
Furthermore, at a given level $k\in\mN_0$, irreducible unitary lowest weight representations only exist for a finite collection of possible lowest weights $h$ and charges $Q$.  They are given by\footnote{The index $c$ should not be confused with the central charge $\mathbf{\bar{c}}$.  Also, for clarity of notation, we will drop the comma in the label $(a,c)$ whenever it is safe to do so.}

\begin{align}
  \begin{matrix}
    h_{ac}=\frac{a(a+2)-c^2}{4\overline{k}}+\frac{[a+c]^2}{8},\\
    Q_{ac}=\frac{c}{2\overline{k}}-\frac{[a+c]}{4},
  \end{matrix}\quad\text{where}\quad
  \begin{matrix}
    a=0,\ldots,k,\\
    \quad |c-[a+c]|\leq a,
  \end{matrix}\label{hq}
\end{align}
where we define $[x]$ to be $0$ if $x$ is even and $1$ if $x$ is odd\footnote{We have actually made a choice here -- choosing $[x]=-1$ for odd $x$ would give an equivalent realisation of the R sector.}.  Here $a+c$ even corresponds to LWRs of the NS sector and $a+c$ odd to LWRs of the R sector.  We will label the indexing set of those $(a,c)$ satisfying $a=0,\ldots,k$ and $|c-[a+c]|\leq a$ at level $k$ by $P_k$.

Di Vecchia et al.~\cite{DiVecchia1986} constructed explicit free fermion representations of each of the possible LWRs via the coset construction of Goddard, Kent and Olive~\cite{Goddard1984}, while an alternative explicit construction using parafermions~\cite{Zamolodchikov1986} was found around the same time by Qiu~\cite{Qiu1987}.  The conjectured characters of these representations\footnote{I am grateful to an anonymous referee for pointing out that the mathematical proof of the correctness of the characters may be incomplete.  As Eholzer and Gaberdiel~\cite{EG1997} point out, the coset construction of Di Vecchia et al~\cite{DiVecchia1986} shows that the $N=2$ SVA at $c<3$ is a subalgebra of the corresponding coset algebra.  The identification of these algebras is equivalent to the identification of the corresponding characters, and in turn follows from the conjectured embedding diagrams.  On the other hand, D\"orrzapf's proof~\cite{Dorrzapf1997} of the correctness of the embedding diagrams relies on the coset identification!  We do not know of any result in the literature that resolves this circular dependency.}
\begin{align*}
  \text{ch}(\tau,z)&=\text{Tr}\left(q^{L_0-\frac{\mathbf{\bar{c}}}{24}}y^{J_o}\right)
\end{align*}
 were calculated shortly afterwards~\cite{Dobrev1987,Dobrev2007,Matsuo1987,Kiritsis1986}\footnote{Embedding diagrams and character formulae were first conjectured in references~\cite{Dobrev1987,Matsuo1987,Kiritsis1986}.  D\"orrzapf~\cite{Dorrzapf1997} later produced more refined embedding diagrams that record the existence of linearly independent uncharged singular vectors at the same level.  See D\"orrzapf~\cite{Dorrzapf1997} and Dobrev~\cite{Dobrev2007} for a discussion.}.  The trace is taken over the states of an irreducible representation of one copy of the SVA, and we use the standard convention that $q=e^{2\pi i\tau},y=e^{2\pi iz}$ for complex parameters $\tau$ and $z$, where $\tau$ is restricted to the upper half complex plane, $\mathcal{H}$.

\subsection{Modular Invariance}\label{mi}
In an SCFT we demand that the bosonic part of the partition function be modular invariant.  Consequently, the objects of interest to us are not the full characters alluded to above, but rather the projections to the bosonic and fermionic states in each irreducible LWR~\cite{Gepner1987}:
\begin{align*}
  \chi_{ac}(\tau,z)&=\text{Tr}_{\mH_{ac}}\left(\frac{1}{2}\left(1+(-1)^{2(J_0-Q_{ac})}\right)q^{L_0-\frac{\mathbf{\bar{c}}}{24}}y^{J_0}\right),\quad (a,c)\in P_k,
\end{align*}
is the trace over the representation $\mH_{ac}$ of the left-hand copy of the SVA and $(-1)^{2(J_0-Q_{ac})}$ is the chiral world-sheet fermion operator.  It is well-defined since, by definition,  $J_0$ has charge $Q_{ac}$ on the lowest weight state $|ac\rangle$ of $\mH_{ac}$, and since the charge of a descendant state differs from $Q_{ac}$ by a half-integer or an integer.  The chiral world-sheet fermion operator commutes with the modes $L_n,J_n$ and anti-commutes with the modes $G^{\pm}_r$, so $\frac{1}{2}(1+(-1)^{2(J_0-Q_{ac})})$ projects to those states created from the lowest weight state $|h_{ac},Q_{ac}\rangle$ by the application of an even number of fermionic modes $G^{\pm}_r$, i.e. states of the form
\begin{align*}
  L_{-n_1}\ldots L_{-n_{\alpha}}J_{-m_1}\ldots J_{-m_{\beta}}G^+_{-l_1}\ldots G^+_{-l_{\gamma}}G^-_{-k_1}\ldots G^-_{k_{\delta}}|h,Q\rangle
\end{align*}
for which $\gamma+\delta$ is even.  Similarly we define
\begin{align*}
  \chi_{k-a,c+\overline{k}}(\tau,z)&=\text{Tr}_{\mH_{ac}}\left(\frac{1}{2}\left(1-(-1)^{2(J_0-Q_{ac})}\right)q^{L_0-\frac{\mathbf{\bar{c}}}{24}}y^{J_0}\right),\quad (a,c)\in P_k,
\end{align*}
the character which counts only those states with $\gamma+\delta$ odd.  The notation $\chi_{k-a,c+\overline{k}}$ is chosen so that the state(s) with the lowest weight that survive the projection have weight $h_{k-a,c+\overline{k}}\mod 1$ and charge $Q_{k-a,c+\overline{k}}\mod 1$ where we have extended the definition of $h$ and $Q$ in equation~\eqref{hq} to the indexing set $Q_k=P_k\cup(\mathbf{j}\cdot P_k)=\{0,\ldots,k\}\times\mZ_{2\overline{k}}$, where $\mathbf{j}$ is the simple current $\mathbf{j}\cdot(a,c)=(k-a,c+\overline{k})$ (see section~\ref{simplecurrents}).

These characters are the building blocks from which we can construct modular invariant partition functions of the minimal models:
\begin{align}
  Z(\tau,z)&=\sum_{\substack{(ac)\in Q_k\\(a'c')\in Q_k}}M_{ac;\,a'c'}\chi_{ac}(\tau,z)\chi_{ac}(\tau,z)^*,\label{Z}
\end{align}
where $M$ is an non-negative integer matrix of multiplicities, and we insist that the vacuum is unique: $M_{00;\,00}=1$.  

The modular group $\text{SL}(2,\mZ)$ acts naturally on $\mathcal{H}\times\mC$ (where $\mathcal{H}$ is the upper half complex plane) via $S:(\tau,z)\mapsto (-\frac{1}{\tau},\frac{z}{\tau})$ and $T:(\tau,z)\mapsto(\tau+1,z)$.  This in turn gives a natural (right) action of $\text{SL}(2,\mZ)$ on the characters $\chi_{ac}$:
\begin{alignat*}{2}
  S\cdot\chi_{ac}(\tau,z)&=\begin{pmatrix}
  0 & -1\\
  1 & 0
  \end{pmatrix}\cdot\chi_{ac}(\tau,z)&&=\chi_{ac}\left(-\frac{1}{\tau},\frac{z}{\tau}\right),\\
  T\cdot\chi_{ac}(\tau,z)&=\begin{pmatrix}
  1 & 1\\
  0 & 1
  \end{pmatrix}\cdot\chi_{ac}(\tau,z)&&=\chi_{ac}(\tau+1,z).
\end{alignat*}
The characters $\{\chi_{ac}\mid (a,c)\in Q_k\}$ transform linearly among themselves under this action and hence span a representation of $\text{SL}(2,\mZ)$.  The $S$- and $T$- matrices are given by
\begin{align}
  S_{ac;\,a'c'}&=2S(k)_{a;a'}S'(2)_{[a+c];[a'+c']}S'(\overline{k})^*_{c;c'},\label{S}\\
  T_{ac;\,a'c'}&=T(k)_{a;a'}T'(2)_{[a+c];[a'+c']}T'(\overline{k})^*_{c;c'}\notag\\
  &=e^{2\pi i(h_{ac}-\frac{\mathbf{\bar{c}}}{24})}\delta_{aa'}\delta_{cc'},\label{T}
\end{align}
 where $S(k)$ is the $S$-matrix of the $\mathfrak{su}(2)$ WZW model~\cite{Knizhnik1984} at level $k$ and $S'(l)$ is the $S$-matrix of the $\mathfrak{u}(1)$ WZW model at level $l$, with similar notation for the $T$-matrices (see e.g.~\cite{Gannon1996} for explicit formulae).  The conformal weight $h_{ac}$ is given in equation~\eqref{hq} and $\delta_{xx'}$ is the Kronecker delta.

Invariance of the partition function $Z(\tau,z)$ in \eqref{Z} under the action of $\text{SL}(2,\mZ)$ is equivalent to
\begin{align}
  M&=SMS^{\dagger},\label{Scommute}\\
  M&=TMT^{\dagger};\notag
\end{align}
or, since $S$ and $T$ are unitary, equivalent to asking that $M$ commutes with both $S$ and $T$.\footnote{We note that this argument relies on the presumed linear independence of the characters $\chi_{ac}$, a fact we have not proven.}

We note here one immediate consequence of modular invariance: using equation~\eqref{T}, we deduce that $T$-invariance is equivalent to
\begin{align}
  M_{ac;\,a'c'}\neq0\quad\Rightarrow\quad h_{ac}-h_{a'c'}\in\mZ.\label{Tcondition}
\end{align}

\subsection{Gannon's Classification}
Gannon's result~\cite{Gannon1996} was to classify all the modular partition functions $Z$ of the form~\eqref{Z} with unique vacuum.  We will refer to the (non-negative integer) matrix of multiplicities $M$ of such a partition function simply as a \emph{modular invariant}\footnote{The terminology \emph{physical invariant} is sometimes employed.  The epithet `physical invariant' is an unfortunate one, particularly as the point of this paper is to determine which of Gannon's modular invariants really are `physical', in the sense of being realised by some full $N=2$ SCFT.  Indeed Gannon~\cite{Gannon2002} has provided examples of possibly `sick' modular invariants for other CFT data that do not correspond to any NIM-rep.}.  We briefly describe how this classification was achieved.

There are two key steps.  The first is to observe that there is a connection between the WZW model $\mathfrak{g}\oplus\mathfrak{h}$ with $\mathfrak{g}=\widehat{\mathfrak{su}(2)}_k\oplus\widehat{\mathfrak{u}(1)}_2$ and $\mathfrak{h}=\widehat{\mathfrak{u}(1)}_{\overline{k}}$, and the minimal models, which can be constructed via the coset representation $\mathfrak{g}/\mathfrak{h}$.  Gannon had already shown~\cite{Gannon1994} that the modular invariants of $\mathfrak{g}/\mathfrak{h}$ could be obtained from the modular invariants of $\mathfrak{g}\oplus\mathfrak{h}$ for various diagonal embeddings of $\mathfrak{h}\subset\mathfrak{g}$ at particular levels.  This phenomenon occurs because of the similarity of the $S$-matrices of the two theories.  In the case of the $N=2$ unitary minimal models we have seen that the $S$-matrix is given by equation~\eqref{S}.  The characters extend naturally to the indexing set $(a,b,c)\in\{0,\ldots,k\}\times\mZ_{4}\times\mZ_{2\overline{k}}=:P'_k$ if we set
\begin{align}
  \chi_{ac}^{(b)}&:=\chi_{ac}\quad\text{when }b=[a+c]\in\{0,1\},\notag\\
  \chi_{k-a,c+\overline{k}}^{(b+2)}&\equiv\chi_{ac}^{(b)}\quad\forall (a,b,c)\in P'_k,\label{newchi}\\
  \chi_{ac}^{(b)}&=0\quad\text{when }a+b+c\not\equiv0\mod2.\notag
\end{align}
With these definitions we find that the characters $\chi_{ac}^{(b)}$ transform under $S$ with $S$-matrix $S(k)\otimes S'(2)\otimes S'(\overline{k})^*$.  Meanwhile the WZW model $\widehat{\mathfrak{su}(2)}_k\oplus\widehat{\mathfrak{u}(1)}_2\oplus\widehat{\mathfrak{u}(1)}_{\overline{k}}$ has characters $\chi_a\chi_b\chi_c$ with $(a,b,c)\in P'_k$, which transform under the action of $S$ with $S$-matrix $S(k)\otimes S'(2)\otimes S'(\overline{k})$.  The crucial observation is that $\chi_a\chi_b\chi_c^*$ transforms under $S$ in exactly the same way as $\chi_{ac}^{(b)}$.  Thus if
\begin{align*}
  \sum M_{abc;\,a'b'c'}\;\chi_{ac}^{(b)}\chi_{a'c'}^{(b')*}
\end{align*}
is a modular invariant of the coset $\mathfrak{g}/\mathfrak{h}$, then
\begin{align*}
  \sum M_{abc';\,a'b'c}\;\chi_a\chi_b\chi_c\chi_{a'}^*\chi_{b'}^*\chi_{c'}^*
\end{align*}
is a modular invariant of the WZW model $\mathfrak{g}\oplus\mathfrak{h}$ (note the interchange of $c$ and $c'$).  This correspondence is injective and thus every $\mathfrak{g}/\mathfrak{h}$ modular invariant is obtained from a $\mathfrak{g}\oplus\mathfrak{h}$ modular invariant, and the subset of $\mathfrak{g}\oplus\mathfrak{h}$ modular invariants corresponding to $\mathfrak{g}/\mathfrak{h}$ modular invariants are precisely those which respect the symmetry in~\eqref{newchi}, i.e. for all $(a,b,c),(a',b',c')\in P'_k$,
\begin{align*}
  M_{k-a,b+2,c;\,a',b',c'+\overline{k}}=M_{a,b,c+\overline{k};\,k-a',b'+2,c'}=M_{abc;\,a'b'c'},\\
  M_{abc;a'b'c'}\neq0\implies a+b+c'\equiv a'+b'+c\equiv 0\mod 2.
\end{align*}
Gannon showed in Lemma 3.1 of~\cite{Gannon1995a} that to check these conditions it is enough to show that the first condition holds on the left- and right-hand vacua:
\begin{align}
  M_{k20;\,00\overline{k}}=M_{00\overline{k};\,k20}=1.\label{Mcondition}
\end{align}
Thus the modular invariant partition functions of the minimal models at level $k$
\begin{align*}
  Z(\tau,z)&=\sum_{\substack{(ac)\in Q_k\\(a'c')\in Q_k}}\widetilde{M}_{ac;\,a'c'}\chi_{ac}(\tau,z)\chi_{a'c'}(\tau,z)^*
\end{align*}
are obtained by
\begin{align*}
  \widetilde{M}_{ac;\,a'c'}=M_{a[a+c]c';\,a'[a'+c']c}
\end{align*}
where $M$ is a modular invariant of $\widehat{\mathfrak{su}(2)}_k\oplus\widehat{\mathfrak{u}(1)}_2\oplus\widehat{\mathfrak{u}(1)}_{\overline{k}}$ satisfying equation~\eqref{Mcondition}, and where, as before, $[x]$ is $0$ or $1$ depending on whether $x$ is even or odd, respectively.

The second step is to classify the modular invariants of $\widehat{\mathfrak{su}(2)}_k\oplus\widehat{\mathfrak{u}(1)}_2\oplus\widehat{\mathfrak{u}(1)}_{\overline{k}}$ subject to equation~\eqref{Mcondition}.  The crucial step is to note that the Verlinde formula~\cite{Verlinde1988} implies that there is a Galois action on the $S$-matrix~\cite{CG93}:
\begin{align*}
  \sigma\cdot S_{abc;\,a'b'c'}=\epsilon_{\sigma}(a,b,c)S_{(abc)^{\sigma};\,a'b'c'}\quad\forall (a,b,c),(a',b',c')\in P'_k
\end{align*}
where $\sigma\in\text{Gal}(K/\mQ)$ for some cyclotomic extension $K$ of $\mQ$, for some $\epsilon\colon P'_k\to\{\pm1\}$ and a permutation $\lambda\mapsto \lambda^{\sigma}$ of $P'_k$.  From this we obtain a selection rule for the modular invariant $M$:
\begin{align*}
  M_{abc;\,a'b'c'}\neq 0\Rightarrow \epsilon_{\sigma}(a,b,c)=\epsilon_{\sigma}(a',b',c').
\end{align*}
This can be solved exactly: we find that either $k\in\{4,8,10,28\}$ or that whenever $M_{000;\,a'b'c'}\neq 0$ we have $a'\in\{0,k\}$.  The former case can be solved by brute force.  The latter solutions comprise the so-called $\mathcal{A}$-$\mathcal{D}$-$\mathcal{E}_7$-invariants\footnote{So called because in the classification of the $\widehat{\mathfrak{su}(2)}_k$ WZW models~\cite{Cappelli1987}, these are precisely the models $\mathcal{A}$,$\mathcal{D}$ and $\mathcal{E}_7$.}\cite{Gannon1995a}.  The $\mathcal{A}$-$\mathcal{D}$-$\mathcal{E}_7$-invariants are defined by the condition
\begin{align*}
  M_{abc;\,000}\neq 0&\Rightarrow (a,b,c)\in\mathcal{J}(0,0,0)\\
  M_{000;\,a'b'c'}\neq 0&\Rightarrow (a',b',c')\in\mathcal{J}(0,0,0)
\end{align*}
where $\mathcal{J}$ is the set of simple currents of the modular invariant $M$~\cite{Intriligator,Schellekens1989a} (see also section~\ref{simplecurrents}).  This is a generalisation of the notion of \emph{simple current invariant}\cite{GatoRivera1992}, a modular invariant $M$ satisfying
\begin{align*}
  M_{abc;\,a'b'c'}\neq 0\quad\Rightarrow\quad (a',b',c')\in \mathcal{J}(a,b,c).
\end{align*}
The classification of the modular invariants of $\widehat{\mathfrak{su}(2)}_k\oplus\widehat{\mathfrak{u}(1)}_2\oplus\widehat{\mathfrak{u}(1)}_{\overline{k}}$ thus reduces to the classification of the $\mathcal{A}$-$\mathcal{D}$-$\mathcal{E}_7$-invariants of $\widehat{\mathfrak{su}(2)}_k\oplus\widehat{\mathfrak{u}(1)}_2\oplus\widehat{\mathfrak{u}(1)}_{\overline{k}}$, which are found using the general method of~\cite{Gannon1995a}.

\subsection{Explicit Classification of Minimal Partition Functions}\label{minimalmodels}
We state the list of partition functions of the minimal models here for two reasons: firstly, it did not appear explicitly in Gannon's paper, and deserves to be accessible in the literature; and secondly because there were a few minor errors in the  application of the main theorem of that paper to the case of $\widehat{\mathfrak{su}(2)}_k\oplus\widehat{\mathfrak{u}(1)}_2\oplus\widehat{\mathfrak{u}(1)}_{\overline{k}}$.  The corrections are highlighted in footnotes.

Throughout this section and the rest of the paper $J$ will denote the $\widehat{\mathfrak{su}(2)}_k$ simple current $J:a\mapsto k-a$.  The notation $\overline{k}$ was defined in equation~\eqref{kbar}.

\medskip

{\bf{$k$ odd:}}
\begin{itemize}
\item We have a modular invariant $\widetilde{M}^0$ for each triple of integers $(v,z,n)$ with $v|\overline{k},\;\overline{k}|v^2$ and $\overline{k}(4z^2-1)/v^2\in\mathbb{Z}$ where $z\in\{1,...,v^2/\overline{k}\}$ and $n\in\{0,1\}$.  Its non-zero entries are
  \begin{align}
    \widetilde{M}^0_{a,c\overline{k}/v;\;a',c'\overline{k}/v}=1\iff
    \left\{\begin{matrix}
    a'=J^{(a+c)n}a\\
    c'\equiv c+(a+c)n\;\text{(mod }2)\\
    c'\equiv 2cz\;\text{(mod }v^2/\overline{k})\\
    \end{matrix}\right\}\label{M^0}
  \end{align}
\end{itemize}

{\bf{$4$ divides $\overline{k}$:}}
\begin{itemize}
\item  We have a modular invariant $\widetilde{M}^{2,0}$ for each triple of integers $(v,z,n)$ with $2v|\overline{k},\;\overline{k}|v^2$ and $y:=\overline{k}(z^2-1)/2v^2\in\mathbb{Z}$ where $z\in\{1,...,2v^2/\overline{k}\}$ and $n\in\{0,1\}$.  Its non-zero entries are
  \begin{align}
    \widetilde{M}^{2,0}_{a,c\overline{k}/v;\;a',c'\overline{k}/v}=1\iff
    \left\{\begin{matrix}
    a'=J^{an+cy}a\\
    c'\equiv cz+ayv^2/\overline{k}\;\text{(mod }2v^2/\overline{k})\\
    \end{matrix}\right\}\label{M^{2,0}}
  \end{align}
  
\item  We have a modular invariant $\widetilde{M}^{2,1}$ for each triple of integers $(v,z,n)$ with $2v|\overline{k}$, $\frac{2v^2}{\overline{k}}\in 2\mathbb{Z}+1$ and $\overline{k}(z^2-1)/2v^2\in\mathbb{Z}$ where $z\in\{1,...,2v^2/\overline{k}\}$ and $n\in\{0,1\}$.  Its non-zero entries are
  \begin{align}
    \widetilde{M}^{2,1}_{a,c\overline{k}/2v;\;a',c'\overline{k}/2v}=1\iff
    \left\{\begin{matrix}
    a\equiv a'\equiv c\equiv c'\;\text{(mod }2)\\
    a'=J^{an+(c+c')/2}a\\
    c'\equiv cz\;\text{(mod }2v^2/\overline{k})
    \end{matrix}\right\}\label{M^{2,1}}
  \end{align}
  
\item  We have a modular invariant $\widetilde{M}^{2,2}$ for each quadruple of integers $(v,z,n,m)$ with $\overline{k}/v$ odd, $v^2/\overline{k}\in \mathbb{Z}$ and $\overline{k}(z^2-1)/4v^2\in\mathbb{Z}$ where $z\in\{1,...,2v^2/\overline{k}\}$ and $n,m\in\{0,1\}$.  Its non-zero entries are
  \begin{align}
    \widetilde{M}^{2,2}_{a,c\overline{k}/v;\;a',c'\overline{k}/v}=1\iff
    \left\{\begin{matrix}
    a'=J^{an+cm}a\\
    c'\equiv cz+(a+c)mv^2/\overline{k}\;\text{(mod }2v^2/\overline{k})
    \end{matrix}\right\}\label{M^{2,2}}
  \end{align}
\end{itemize}

{\bf{$4$ divides $k$}}
\begin{itemize}
\item  If $8|k+4$ then we have a modular invariant $\widetilde{M}^{4,0}$ for each quadruple of integers $(v,z,n,m)$ with $\overline{k}/2v\in\mathbb{Z},\;x:=(1/4+v^2/2\overline{k})\in\mathbb{Z}$ and $\overline{k}(z^2-1)/2v^2\in\mathbb{Z}$ where $z\in\{1,...,2v^2/\overline{k}\}$ and $m,n\in\{0,1\}$.  Its non-zero entries are
  \begin{align}
    \widetilde{M}^{4,0}_{a,c\overline{k}/2v;\;a',c'\overline{k}/2v}=1\!\!\iff\!\!
    \left\{\begin{matrix}
    c+c'\equiv a\equiv a'\;\text{(mod }2)\\
    a'=J^{ax+cn+c(1-c)/2}a\\
    c'\equiv cz\;\text{(mod }2v^2/\overline{k})\\
    2c'm+c'(1-c')\equiv 2cn+c(1-c)\;\text{(mod }4)
    \end{matrix}\right\}\label{M^{4,0}}
  \end{align}
  Note that $\widetilde{M}^{4,0}$ is only symmetric when $m=n$.  In fact $(\widetilde{M}(v,z,n,m))^T=\widetilde{M}(v,z,m,n)$.  Note also that the condition that $x$ be an integer follows directly from the conditions that $8|k+4$ and $\overline{k}|2v^2$.
  
\item  If $8|k$ then we have a modular invariant\footnote{In the original classification the modulo $8$ condition was only given modulo $1$} $\widetilde{M}^{4,1}$ for each quadruple of integers $(v,z,x,y)$ with $v|\overline{k},\;\overline{k}|v^2,\;2\overline{k}(4z^2-1)/v^2\equiv 7\;\text{(mod }8)$ where $z\in\{1,...,v^2/\overline{k}\}$ and $x,y\in\{1,3\}$.  Its non-zero entries are
  \begin{align}
    \widetilde{M}^{4,1}_{a,c\overline{k}/v;\;a',c'\overline{k}/v}=1+\delta_{a,k/2}\iff
    \left\{\begin{matrix}
    a \equiv a'\equiv 0\;\text{(mod }2)\\
    a'=J^la\;\;\;\text{for some }l\in\mathbb{Z}\\
    c'\equiv 2cz\;\text{(mod }v^2/2\overline{k})\\
    c(c-x)\equiv 2c'z\;\text{(mod }4)\\
    c'(c'-y)\equiv 2cz\;\text{(mod }4)
    \end{matrix}\right\}\label{M^{4,1}}
  \end{align}
  Note that $\widetilde{M}^{4,1}$ is only symmetric when $x=y$.  In fact $(M(v,z,x,y))^T=M(v,z,y,x)$.  Note also that the condition $2\overline{k}(4z^2-1)/v^2\equiv 7\;\text{(mod }8)$ is equivalent to $2\overline{k}(4z^2-1)/v^2\in\mathbb{Z}$ and $k/8\equiv z\;\text{(mod }2)$.
  
\item  We have a modular invariant $\widetilde{M}^{4,2}$ for each triple of integers $(v,z,x)$ with $2v|\overline{k}$, $\overline{k}|2v^2$ and $\overline{k}(z^2-1)/2v^2\in\mathbb{Z}$ where $z\in\{1,...,2v^2/\overline{k}\}$ and $x\in\{1,3\}$.  Its non-zero entries are
  \begin{align}
    \widetilde{M}^{4,2}_{a,c\overline{k}/2v;\;a',c'\overline{k}/2v}=1+\delta_{a,k/2}\iff
    \left\{\begin{matrix}
    a \equiv a'\equiv 0\;\text{(mod }2)\\
    a'=J^la\;\;\;\text{for some }l\in\mathbb{Z}\\
    c'\equiv cz\;\text{(mod }2v^2/\overline{k})\\
    c'\equiv cx\;\text{(mod }4)
    \end{matrix}\right\}\label{M^{4,2}}
  \end{align}
  
\item  We have a modular invariant\footnote{In the original classification of the $\widehat{\mathfrak{su}(2)}_k\oplus\widehat{\mathfrak{u}(1)}_2\oplus\widehat{\mathfrak{u}(1)}_{\overline{k}}$ invariants, the non-zero entries of $M^{4,3}$ should have read $M_{abc;\,J^la,bx+2l,cv+2lv}=1$ with $(c+bv-av)v/\overline{k}\in\mZ$ and $l\in\mZ$, and $z$ should be allowed to run from $1$ to $8v^2/\overline{k}$ rather than only up to $4v^2/\overline{k}$.} $\widetilde{M}^{4,3}$ for each triple $(v,z,n)$ with $2v|\overline{k}$, $\overline{k}|2v^2$ and $\overline{k}(z^2-1)/4v^2\in\mathbb{Z}$ where $z\in\{1,...,8v^2/\overline{k}\}$ and $n\in\{0,1\}$.  Its non-zero entries are
  \begin{align}
    \widetilde{M}^{4,3}_{a,c\overline{k}/2v;\;a',c'\overline{k}/2v}=1\iff
    \left\{\begin{matrix}
    a'=J^{(a+c)n}a\\
    c'\equiv cz\;\text{(mod }2v^2/\overline{k})\\
    c'\equiv cz+2(a+c)n\;\text{(mod }4)\\
    \end{matrix}\right\}\label{M^{4,3}}
  \end{align}
\end{itemize}

{\bf{Exceptional Invariants}}

Throughout this section we will write $A^k,D^k$ and $E^k$ for the $\widehat{\mathfrak{su}(2)}_k$ modular invariants at $k=10,16$ and $28$.
\begin{itemize}
\item  When $k=10$ we have a modular invariant $\widetilde{E}^{10}_1$ for the 2 pairs of integers $(v,z)$ with $v=6$ and $z\in\{1,5\}$.  Note that for these values of $k,v$ and $z$ the modular invariant $\widetilde{M}^{2,0}$ factors into $N\otimes M$, since the parameter $y$ turns out to be even, where $N$ equals either $A^{10}$ or $D^{10}$ and $M$ is the $\widehat{\mathfrak{u}(1)}$ part.  Then $\widetilde{E}^{10}_1=E^{10}\otimes M$.  The non-zero entries of $M$ are
  \begin{align}
    M_{2c;\;2c'}=1\iff
    \left\{\begin{matrix}
    c'\equiv cz\;\text{(mod }6)\\
    \end{matrix}\right\}\label{E^{10}_1}
  \end{align}
  
\item  When $k=10$ we have a modular invariant $\widetilde{E}^{10}_2$ for the 8 quadruples $(v=12,z,n=0,m)$ with $z\in\{1,7,17,23\}$ and $m\in\{0,1\}$.  Then $\widetilde{E}^{10}_2$ is given by
  \begin{align}
    (\widetilde{E}^{10}_2)_{ac;\;a'c'}=1\iff
    \left\{\begin{matrix}
    E^{10}_{J^{cm}a;a'}=1\\
    c'\equiv cz+12(a+c)m\;\text{(mod }24)\\
    \end{matrix}\right\}\label{E^{10}_2}
  \end{align}
  
\item  When $k=16$ we have a modular invariant $\widetilde{E}^{16}_1$ for the 12 quadruples of integers $(v,z,x,y)$ with either $v=6,z=2$ or $v=18,z\in\{4,5\}$, and $x,y\in\{1,3\}$.  Note that the modular invariant $\widetilde{M}^{4,1}$ factors into $D^{16}\otimes M$ where $M$ is the $\widehat{\mathfrak{u}(1)}$ part.  Then $\widetilde{E}^{16}_1=E^{16}\otimes M$.  The non-zero entries of $M$ are
  \begin{align}
    M_{18c/v;\;18c'/v}=1\iff
    \left\{\begin{matrix}
    c'\equiv 2cz\;\text{(mod }v^2/36)\\
    c(c-x)\equiv 0\;\text{(mod }4)\\
    c'(c'-y)\equiv 0\;\text{(mod }4)
    \end{matrix}\right\}
  \end{align}
  
\item  When $k=16$ we have a modular invariant $\widetilde{E}^{16}_2$ for the 6 triples $(v,z,x)$ with either $v=3,z=1$ or $v=9,z\in\{1,8\}$, and $x\in\{1,3\}$.  Note that $\widetilde{M}^{4,2}$ factors into $D^{16}\otimes M$ where $M$ is the $\widehat{\mathfrak{u}(1)}$ part.  Then $\widetilde{E}^{16}_2=E^{16}\otimes M$.  The non-zero entries of $M$ are:
  \begin{align}
    M_{9c/v;\;9c'/v}=1\iff
    \left\{\begin{matrix}
    c'\equiv cz\;\text{(mod }v^2/9)\\
    c'\equiv cx\;\text{(mod }4)
    \end{matrix}\right\}
  \end{align}
  
\item  When $k=28$ we have a modular invariant\footnote{There are 16 modular invariants described as coming from $M^{4,0}$ in the original classification, but no such invariants in fact exist.} $\widetilde{E}^{28}$ for the 8 triples $(v=15,z,x)$ with $z\in\{1,4,11,14\}$ and $x\in\{1,3\}$.  Note that $\widetilde{M}^{4,2}$ factors into $D^{28}\otimes M$ where $M$ is the $\widehat{\mathfrak{u}(1)}$ part.  Then $\widetilde{E}^{28}=E^{28}\otimes M$.  The non-zero entries of $M$ are
  \begin{align}
    M_{c;\;c'}=1\iff
    \left\{\begin{matrix}
    c'\equiv cz\;\text{(mod }15)\\
    c'\equiv cx\;\text{(mod }4)
    \end{matrix}\right\}\label{E^{28}}
  \end{align}
\end{itemize}

Those modular invariants corresponding to the space-time supersymmetric theories of Cecotti and Vafa will be identified in section~\ref{spacetime}.

\subsection{Simple Examples}
To illustrate the foregoing classification, and to demonstrate that, at least for the lowest levels, the partition functions turn out to be given in terms of familiar functions, we will calculate the partition functions explicitly for levels $k=1$ and $k=2$.

\subsubsection{\texorpdfstring{$k=1$}{k=1}}\label{k=1}
Level $k=1$ yields $N=2$ superconformal unitary minimal models with central charge $\mathbf{\bar{c}}=1$.  We can express the characters in terms of familiar functions:
\begin{align*}
  \chi_{ac}(\tau,z)&=K^{(6)}_{2c-3[a+c]}(\tau,z)
\end{align*}
where $K_x^{(6)}$ are the $\widehat{\mathfrak{u}(1)}_6$ characters\footnote{The Kac-Moody algebra $\widehat{\mathfrak{u}(1)}$ does not have levels as such, since the generators can always be rescaled.  We borrowed the notation $\widehat{\mathfrak{u}(1)}_l$ from~\cite{DiFrancesco1997}.} defined by
\begin{align}\label{Keq}
K^{(l)}_x(\tau,z)&=\frac{1}{\eta(\tau)}\sum_{Q\in\Gamma^{(l)}_x}q^{lQ^2}y^{Q},\;\;\;x\in\mathbb{Z}_{2l},
\end{align}
the shifted lattice $\Gamma^{(l)}_x$ is given by $\Gamma^{(l)}_x=\left\{\left.\left(n+\frac{x}{2l}\right)\right|n\in\mathbb{Z}\right\}$ and $\eta$ is the Dedekind $\eta$-function.  We can then read off from section~\ref{minimalmodels} the partition functions of the $4$ minimal models with $c=1$.  We label the four partition functions by the parameters $(v,z,n)$ (see equation~\eqref{M^0} for notation):
\begin{align*}
  Z(3,2,0)(\tau,z)&=Z_{R=\sqrt{6}}(\tau,z);\\
  Z(3,1,1)(\tau,z)&=Z_{R=\frac{1}{\sqrt{6}}}(\tau,z),\\
  Z(3,2,1)(\tau,z)&=Z_{R=\sqrt{\frac{3}{2}}}(\tau,z);\\
  Z(3,1,0)(\tau,z)&=Z_{R=\sqrt{\frac{2}{3}}}(\tau,z);
\end{align*}
where $Z_R$ is the partition function of the boson on the circle at radius $R$ (see e.g~\cite{Ginsparg1988}):\footnote{In our normalisation the self-dual radius is $R=1$.  Some authors use $R=\sqrt{2}$.}
\begin{align}
  Z_{R}(\tau,z)&=\frac{1}{\left|\eta(\tau)\right|^2}\sum_{(Q,\overline{Q})\in\Gamma_R}q^{lQ^2}y^{Q}\overline{q}^{l\overline{Q}^2}\overline{y}^{\overline{Q}},\label{Z_R}\\
\Gamma_{R}&=\left\{\frac{1}{2\sqrt{l}}\left(\left.\frac{n}{R}+mR,\;\frac{n}{R}-mR\right)\right| n,m\in\mathbb{Z}\right\},\notag
\end{align}
where here $l=6$.  The pair $(Q,\overline{Q})\in\Gamma_R$ labels a conformal primary state with $U(1)$ charges $(Q,\overline{Q})$ and conformal weights $(h,\overline{h})=(6Q^2,6\overline{Q}^2)$.\footnote{It is perhaps more usual to re-scale the $U(1)$ current for the boson on the circle by $\sqrt{12}$ to obtain $h=\frac{Q^2}{2}$.  The price, of course, is that the $N=2$ algebra, which is a symmetry of these $c=1$ theories at the special radii $R,R^{-1}\in\{\sqrt{6},\sqrt{\frac{3}{2}}\}$,  will then differ from its usual form: e.g. we would find $[J_0,G^{\pm}_{r}]=\pm\sqrt{3}G^{\pm}_r$.  See Waterson~\cite{Waterson1986} for an explicit construction of the irreducible representations of the unitary $N=2$ minimal models at $c=1$.}

The partition function with $(z,v,n)=(3,2,0)$ is that of the diagonal model.  The first and second partition functions, and the third and fourth partition functions belong to mirror symmetry pairs.  Mirror symmetry is realised by acting by the charge conjugation matrix $C=S^2$ on one of the chiral sectors.  At the level of primary states, mirror symmetry acting on the left-hand representations maps states with $U(1)$ charges $(Q,\overline{Q})$ to states with charges $(-Q,\overline{Q})$.  This implies that one model can be obtained from the other by relabelling the generators of the left $U(1)$ current:
\begin{align*}
  \{L_n,J_n,G^{\pm}_r,\overline{L}_n,\overline{J}_n,\overline{G}^{\pm}_r\}\rightarrow\{L_n,-J_n,G^{\mp}_r,\overline{L}_n,\overline{J}_n,\overline{G}^{\pm}_r\}.
\end{align*}
Thus the two mirror symmetric models describe identical physics, and we would normally consider them to be equivalent theories.  However, since they give rise to different partition functions, it will be convenient to treat them as belonging to separate theories.  The analogue is true for mirror symmetry acting on the right-hand states.

We note that combining both left- and right- mirror symmetry transformations yields the \emph{charge conjugation} transformation\footnote{We emphasise that acting with the charge conjugation matrix $C$ on one chiral halve yields the mirror symmetry transformation; acting on both halves simultaneously yields the charge conjugation transformation.}, which acts on charges of states via $(Q,\overline{Q})\rightarrow(-Q,-\overline{Q})$.  Since the charge conjugation matrix $C$ satisfies $C^2=S^4=\text{Id}$, we see that this leaves the partition functions invariant.  We will therefore consider charge conjugate theories to be identical.

In the current case, we see that mirror symmetry coincides with the $T$-duality~\cite{Buscher1987,Rocek1991} transformation, which interchanges $Z_R$ and $Z_{\frac{1}{R}}$.

\subsubsection{\texorpdfstring{$k=2$}{k=2}}\label{k=2}
The level $k=2$ models correspond to the $N=2$ superconformal unitary minimal models with central charge $\mathbf{\bar{c}}=\frac{3}{2}$.  Again, we can express the characters in terms of familiar functions:
\begin{align*}
  \chi_{ac}(\tau,z)&=\eta(\tau)\,c_{a,c-[a+c]}^{(2)}(\tau)\,K^{(4)}_{c-2[a+c]}(\tau,z)
\end{align*}
where $K_x^{(4)},x\in\mZ_8$ are the $\widehat{\mathfrak{u}(1)}_l$ characters given in equation~\eqref{Keq} for $l=4$ and $c^{(2)}_{a,c}$ are the level $2$ $\mathfrak{su}(2)$ string functions (see e.g.~\cite{Kac1984}).  The string functions can be written in terms of the Jacobi theta functions and the Dedekind eta function as follows:
\begin{align*}
  \eta(\tau)c_{a,c}^{(2)}(\tau)&=\left\{\begin{matrix}
  \sqrt{\frac{\theta_2(\tau,0)}{2\eta(\tau)}}&\text{if }a=1\\
  \frac{1}{2}\left(\sqrt{\frac{\theta_3(\tau,0)}{\eta(\tau)}}+(-1)^{\frac{a+c}{2}}\sqrt{\frac{\theta_4(\tau,0)}{\eta(\tau)}}\right)&\text{if $a$ is even}.
  \end{matrix}
  \right.
\end{align*}
We can now evaluate the five modular invariant partition functions\footnote{When we count the number of simple current invariants in theorem~\ref{countingtheorem}, we will see that our formula predicts ten partition functions at level 2.  This discrepancy arises from the identity $\mathcal{A}_2=\mathcal{D}_2$, which does not generalise to other levels $k$.  Thus we only expect to find five theories at level $k=2$.} using the labels $(0;v,z)$ for the unique $\widetilde{M}^{2,0}$ invariant (see equation~\eqref{M^{2,0}} -- we have dropped the label $n$ since $n=0$ or $1$ give the same partition function for $k=2$) and labels $(2;v,z,m)$ for the four partition functions in the family $\widetilde{M}^{2,2}$ (see equation~\eqref{M^{2,2}}--again we have dropped the $n$ label).
\begin{align*}
  Z(0;2,1)(\tau,z)&=Z_{\text{Ising}}(\tau)Z_{R=1}(\tau,z);\\
  Z(2;4,1,0)(\tau,z)&=Z_{\text{Ising}}(\tau)Z_{R=2}(\tau,z);\\
  Z(2;4,7,1)(\tau,z)&=Z_{\text{Ising}}(\tau)Z_{R=\frac{1}{2}}(\tau,z);\\
  Z(2;4,7,0)(\tau,z)&=\frac{1}{2}\!\sum_{c\in\mathbb{Z}_8}\!\!\left(\left|\frac{\theta_3(\tau,0)}{\eta(\tau)}\right|+(-1)^c\left|\frac{\theta_4(\tau,0)}{\eta(\tau)}\right|\right)K^{(4)}_c(\tau,z)K^{(4)}_{3c}(\tau,z)^*\\
  &\quad+\frac{1}{2}\left|\frac{\theta_2(\tau,0)}{\eta(\tau)}\right|\sum_{c\in\mathbb{Z}_8}K^{(4)}_c(\tau,z)K^{(4)}_{3c+4}(\tau,z)^*;\\
  Z(2;4,1,1)(\tau,z)&=\frac{1}{2}\!\sum_{c\in\mathbb{Z}_8}\!\!\left(\left|\frac{\theta_3(\tau,0)}{\eta(\tau)}\right|+(-1)^c\left|\frac{\theta_4(\tau,0)}{\eta(\tau)}\right|\right)K^{(4)}_c(\tau,z)K^{(4)}_{5c}(\tau,z)^*\\
  &\quad+\frac{1}{2}\left|\frac{\theta_2(\tau,0)}{\eta(\tau)}\right|\sum_{c\in\mathbb{Z}_8}K^{(4)}_c(\tau,z)K^{(4)}_{5c+4}(\tau,z)^*,
\end{align*}
where here
\begin{align*}
  Z_{\text{Ising}}&=\frac{1}{2}\left(\left|\frac{\theta_2(\tau,0)}{\eta(\tau)}\right|+\left|\frac{\theta_3(\tau,0)}{\eta(\tau)}\right|+\left|\frac{\theta_4(\tau,0)}{\eta(\tau)}\right|\right)
\end{align*}
is the partition function of the Ising model (see e.g.~\cite{Ginsparg1988}), and $Z_R$ is the partition function of the boson on the circle given in equation~\eqref{Z_R} with $l=4$.

We note that the second partition function is that of the diagonal model.  The first partition function belongs to a self-mirror-symmetric model, and the second and third, and the fourth and fifth partition functions belong to mirror symmetry pairs.  On the level of primary states mirror symmetry acts on the left-hand representations by mapping the primary state $|\text{Ising}\rangle\otimes|Q,\overline{Q}\rangle$ to $|\text{Ising}\rangle\otimes|-Q,\overline{Q}\rangle$, and similarly on the right-hand representations.  This induces the transformation $K_c\mapsto K_{-c}$ on the $U(1)$ characters.  On the self-mirror-symmetric model and the first mirror symmetry pair, mirror symmetry is realised via $T$-duality, by interchanging $Z_{R}$ and $Z_{\frac{1}{R}}$.

\section{Consequences of Gannon's classification}
\subsection{Classification of Theories with Space-Time Supersymmetry}\label{spacetime}
In this section we show that those partition functions belonging to space-time supersymmetric models fall into the well-known \cADE pattern in accordance with~\cite{Cecotti1993}.  Specifically we will find those partition functions that satisfy the following condition:  the R$\otimes$R sector of the theory is obtained from NS$\otimes$NS sector under simultaneous spectral flow by half a unit on both chiral halves of the theory, and the NS$\otimes$R and R$\otimes$NS sectors are similarly interchanged.  The spectral flow is rather easy to describe in our notation: it simply maps between the NS sector and the R sector via $(a,c)\leftrightarrow(a,c+1)$ where $a+c$ is even.  One can check using equations~\eqref{c} and~\eqref{hq} that for $a+c$ even we have
\begin{align*}
  h_{ac}\rightarrow h_{a,c+1}=h_{ac}-Q_{ac}+\frac{\mathbf{\bar{c}}}{24},
\end{align*}
as expected from e.g.~\cite{Greene1996}.  The constraint that a theory should be invariant under the interchange of NS$\otimes$NS$\leftrightarrow$R$\otimes$R and NS$\otimes$R$\leftrightarrow$R$\otimes$NS is a very strong one.  In particular, since the vacuum representation must be present in any theory, the representation obtained from the vacuum by spectral flow should be present in the R$\otimes$R sector; i.e. $M_{01;\,01}\neq0$.  One can read off from the explicit list in section~\ref{minimalmodels} that the only space-time supersymmetric theories have the following partition functions:
\begin{align*}
  \widetilde{M}^0(v=\overline{k},2z=1,n=0)=\mathcal{A}_k\otimes I_{2\overline{k}},&&k\text{ odd}\\
  \widetilde{M}^{2,2}(v=\overline{k},z=1,n=0,m=0)=\mathcal{A}_k\otimes I_{2\overline{k}},&&4\text{ divides }\overline{k}\\
  \widetilde{M}^{2,2}(v=\overline{k},z=1,n=1,m=0)=\mathcal{D}_k\otimes I_{2\overline{k}},&&4\text{ divides }\overline{k}\\
  \widetilde{M}^{4,3}(v=\frac{\overline{k}}{2},z=1,n=0)=\mathcal{A}_k\otimes I_{2\overline{k}},&&4\text{ divides }k\\
  \widetilde{M}^{4,2}(v=\frac{\overline{k}}{2},z=1,x=1)=\mathcal{D}_k\otimes I_{2\overline{k}},&&4\text{ divides }k\\
  \widetilde{E}^{10}_2(v=12,z=1,n=0,m=0)=\mathcal{E}_{10}\otimes I_{2\overline{k}},&&k=10\\
  \widetilde{E}^{16}_2(v=9,z=1,x=1)=\mathcal{E}_{16}\otimes I_{2\overline{k}},&&k=16\\
  \widetilde{E}^{28}(v=15,z=1,x=1)=\mathcal{E}_{28}\otimes I_{2\overline{k}},&&k=28
\end{align*}
Here the $\mathcal{A}_k$,$\mathcal{D}_k$,$\mathcal{E}_k$ are the $\widehat{\mathfrak{su}(2)}_k$ modular invariants of~\cite{Cappelli1987} and the $I_{2\overline{k}}$ are $\widehat{\mathfrak{u}(1)}_{\overline{k}}$ diagonal invariants\footnote{We use the notation $I_{2\overline{k}}$ since they are $2\overline{k}\times2\overline{k}$ matrices.  Some authors use $I_{\overline{k}}$.}.
These theories have no NS$\otimes$R or R$\otimes$NS sectors, and the NS$\otimes$NS sector can be recovered from the R$\otimes$R sector via spectral flow by half a unit in the opposite direction.  

The familiar \cADE pattern has emerged.  It is quite remarkable that the \cADE classification arises already at the level of partition functions.

We note here that there is (at least) one space-time supersymmetric minimal model in each ``orbifold class'' of the unitary $N=2$ minimal models; that is, every partition function in Gannon's list can be mapped to one of the space-time supersymmetric partition functions by an orbifolding constructed in section~\ref{orbifold}.

\subsection{Simple Currents and Fusion Rules}\label{simplecurrents}

In the study of conformal field theories, a rich symmetry structure arises out of the so-called \emph{simple currents}~\cite{Intriligator,Schellekens1989a,Schellekens1989b}.  A simple current is a primary field which upon fusion with any other field yields precisely one primary field (plus its descendants).  The simple currents can therefore be found from the fusion coefficients $N_{aa'}^{a''}$ defined by
\begin{align*}
  [\phi_a]\times[\phi_{a'}]=\sum_{a''\in P}N_{aa'}^{a''}[\phi_{a''}]
\end{align*}
where $\phi_a$ are primary fields labelled by some indexing set $P$.  $[\phi_a]$ represents a sum over the primary field $\phi_a$ and its descendants.  $N_{aa'}^{a''}$ counts the multiplicity of the field $\phi_{a''}$ appearing in the OPE of $\phi_a$ and $\phi_{a'}$.

\subsubsection{The Verlinde Formula and Fusion Rules}\label{verlindesection}
The Verlinde formula~\cite{Verlinde1988} gives a surprising and elegant expression for fusion rules in terms of the $S$-matrix for (bosonic) rational CFTs\cite{MS1988}.  Inspired by this we define (for the $S$-matrix of the unitary $N=2$ minimal models given in equation~\eqref{S})
\begin{align}\label{Verlinde}
  N_{ac;\,a'c'}^{a''c''}:=\sum_{(d,f)\in Q_k}\frac{S_{ac;df}S_{a'c';df}S_{a''c'';df}^*}{S_{00;df}}.
\end{align}
We want to interpret $N_{ac,a'c'}^{a''c''}$ as the fusion coefficients for the $N=2$ minimal models.  We will return to make a case for this claim after the next lemma, in which we show that the numbers $N_{ac,a'c'}^{a''c''}$ are integers, and are in fact related to the fusion coefficients of familiar bosonic CFTs.
\begin{lemma}\label{fusionrules}
  Fix $(ac),(a'c'),(a'',c'')\in Q_k$.  We have
  \begin{align}\label{N=2fusion}
    N_{ac;\,a'c'}^{a''c''}=
    \left\{\begin{matrix}
    \left(N^{\widehat{\mathfrak{su}(2)}_k}\right)^{a''}_{aa'}\left(N^{\widehat{\mathfrak{u}(1)}_{\overline{k}}}\right)^{c''}_{cc'}, & \text{if  }[a+c][a'+c']=0\\
    \left(N^{\widehat{\mathfrak{su}(2)}_k}\right)^{k-a''}_{aa'}\left(N^{\widehat{\mathfrak{u}(1)}_{\overline{k}}}\right)^{c''+\overline{k}}_{cc'}, & \text{if  }[a+c][a'+c']=1\\
    \end{matrix}\right\}
  \end{align}
\end{lemma}
  Here $N^{\widehat{\mathfrak{su}(2)}_k}$ and $N^{\widehat{\mathfrak{u}(1)}_{\overline{k}}}$ are the fusion coefficients for the WZW models~\cite{Wess1971,Witten1983} obtained from $\mathfrak{su}(2)$ at level $k$~\cite{Gepner1986a} and $\widehat{\mathfrak{u}(1)}$ at level $\overline{k}$ respectively.  In general, they read
  \begin{align*}
    \left(N^{\widehat{\mathfrak{su}(2)}_k}\right)^{a''}_{aa'}&=\delta(|a-a'|\leq a''\!\!\leq \min(a+a',2k-a-a'))\;\delta(a+a'\equiv a''\!\!\!\!\!\mod2)\\
    \left(N^{\widehat{\mathfrak{u}(1)}_{l}}\right)^{c''}_{cc'}&=\delta(c+c'\equiv c''\!\!\!\mod 2l),
  \end{align*}
where $\delta(\text{condition})=1$ if `condition' is satisfied, and $0$ otherwise.  In particular, we see that $N_{ac;\,a'c'}^{a''c''}$ is only non-zero if $a+c+a'+c'+a''+c''\equiv0\mod2$.  So if we can interpret the $N$ as fusion coefficients of the minimal models then we obtain the following selection rules for the NS ($[a+c]=0$) and R ($[a+c]=1$) sectors:
\begin{align*}
  NS\times NS&\sim NS & NS\times R&\sim R\\
  R\times NS&\sim R & R\times R&\sim NS.
\end{align*}
\begin{proof}
  It is possible to expand the expression~\eqref{Verlinde} into a sum of products of sines and exponentials which can be simplified at great tedium.  We present here a very simple proof using simple currents of the $S$-matrices of the WZW models obtained from $\mathfrak{su}(2)$ and $\mathfrak{u}(1)$.  Simple currents are explained in detail in the following section, but for now we will just use the fact that
  \begin{align}\label{scurrents}
    \begin{split}
      S_{k-a;a'}&=(-1)^{a'}S_{a;a'},\quad a,a'\in\{0,\ldots,k\}\\
      S'_{b+2;b'}&=(-1)^{b'}S_{b;b'},\quad b,b'\in\mZ_{4},\\
      S''_{c+\overline{k};c'}&=(-1)^{c'}S'_{c;c'},\quad c,c'\in\mZ_{2\overline{k}},
    \end{split}
  \end{align}
  where for brevity we have written $S$ for the $\mathfrak{su}(2)$ $S$-matrix at level $k$, and $S'$ and $S''$ for the $\mathfrak{u}(1)$ $S$-matrix at levels $2$ and $2\overline{k}$ respectively.
  
  For the rest of the proof, let us slightly abuse notation by denoting by $J$ the permutations $a\mapsto k-a,b\mapsto b+2$ and $c\mapsto c+\overline{k}$, as well as the permutation of $P_k'$ given by $J(a,b,c)=(Ja,Jb,Jc)$ (cf equation~\eqref{newchi}).
  
  The author is grateful to an anonymous referee for pointing out that the right-hand side of the $N=2$ fusion rules~\eqref{N=2fusion} can be neatly expressed as
  \begin{align*}
    \text{RHS}&=\sum_{j=0}^1\left(N^{\text{WZW}}\right)_{abc;a'b'c'}^{J^j(a''b''c'')},
  \end{align*}
  where $N^{\text{WZW}}=N^{\widehat{\mathfrak{su}(2)}_k}\otimes N^{\widehat{\mathfrak{u}(1)}_2}\otimes N^{\widehat{\mathfrak{u}(1)}_{\overline{k}}}$ are the fusion coefficients of the WZW model $\widehat{\mathfrak{su}(2)}_k\oplus\widehat{\mathfrak{u}(1)}_2\oplus\widehat{\mathfrak{u}(1)}_{\overline{k}}$, and we have written $b=[a+c],b'=[a'+c']$ and $b''=[a''+c'']$.  This observation allows for a more elegant proof, which also shows how lemma~\ref{fusionrules} would generalise for certain other cosets.

  We start by expanding the fusion rules using the Verlinde formula for the WZW model $\widehat{\mathfrak{su}(2)}_k\oplus\widehat{\mathfrak{u}(1)}_2\oplus\widehat{\mathfrak{u}(1)}_{\overline{k}}$ in terms of the $S$-matrix of the WZW theory $\tilde{S}=S\otimes S'\otimes S''$:
  \begin{align*}
    \sum_{j=0}^1\left(N^{\text{WZW}}\right)_{abc;a'b'c'}^{J^j(a''b''c'')}&=\sum_{j=0}^1\sum_{(def)\in P'_k}\frac{\tilde{S}_{abc;def}\tilde{S}_{a'b'c';def}\tilde{S}^*_{J^j(a''b''c'');def}}{\tilde{S}_{000;def}}\\
    &=2\sum_{\substack{(def)\in P'_k\\d+e+f\, \text{even}}}\frac{\tilde{S}_{abc;def}\tilde{S}_{a'b'c';def}\tilde{S}^*_{a''b''c'';def}}{\tilde{S}_{000;def}}
    \intertext{where we used equation~\eqref{scurrents} in the last line.  Writing $P''_k$ for those $(def)\in P'_k$ with $e=[d+f]$ we have}
    &=2\sum_{j=0}^1\sum_{(def)\in P''_k}\frac{\tilde{S}_{abc;J^j(def)}\tilde{S}_{a'b'c';J^j(def)}\tilde{S}^*_{a''b''c'';J^j(def)}}{\tilde{S}_{000;J^j(def)}}\\
    &=4\sum_{(def)\in P''_k}\frac{\tilde{S}_{abc;def}\tilde{S}_{a'b'c';def}\tilde{S}^*_{a''b''c'';def}}{\tilde{S}_{000;def}}\\
    &=4\sum_{(def)\in P''_k}\frac{\tilde{S}_{abc;d,e,-f}\tilde{S}_{a'b'c';d,e,-f}\tilde{S}^*_{a''b''c'';d,e,-f}}{\tilde{S}_{000;d,e,-f}}
  \end{align*}
  where in the second line we used equations~\eqref{scurrents} again along with the fact that $b=[a+c], b'=[a'+c']$ and $b''=[a''+c'']$; and in the last line we used $(d,e,f)\in P''_k\iff(d,e,-f)\in P''_k$.  Finally, we note that $f\mapsto -f$ implements charge conjugation in the $\mathfrak{u}(1)$ WZW model: $S''_{c;-f}=S''^*_{c;f}$.  Thus we can relate the WZW $S$-matrix to the coset $S$-matrix by equation~\eqref{S}:
  \begin{align*}
    2\tilde{S}_{abc;d,e,-f}&=2S_{ad}S'_{be}S''^*_{c,f}=S_{ac;df},
  \end{align*}
  and we arrive at
  \begin{align*}
    \sum_{j=0}^1\left(N^{\text{WZW}}\right)_{abc;a'b'c'}^{J^j(a''b''c'')}&=\sum_{(df)\in Q_k}\frac{S_{ac;df}S_{a'c';df}S^*_{a''c'';df}}{S_{00;df}}\\
    &=N_{ac;a'c'}^{a''c''}
  \end{align*}
  as required.\qed
\end{proof}
Recall that the fusion in a (bosonic) CFT describes how the different conformal families combine under the operator product expansion (OPE).  Let $\phi_a(z),\phi_b(w)$ be primary fields (where we consider only the holomorphic part dependent on $z,w$).  Then the fusion of $\phi_a(z)$ with $\phi_b(w)$ is given by
\begin{align}\label{conformalfusion}
  \phi_a(z)\phi_b(w)&=\sum_{c\in P}C_{ab}^c(z-w)^{h_c-h_a-h_b}\left[\phi_c(w)+\sum_{n>0}(z-w)^n\phi_c^{(n)}(w)\right]
\end{align}
where $C_{ab}^c\in\mC$ (which when multiplied by their anti-holomorphic counterpart $C_{\bar{a}\bar{b}}^{\bar{c}}$ yield the \emph{OPE coefficients}), $h_x\in\mC$ is the conformal weight of the primary field $\phi_x$ and $P$ labels the set of primary fields.  $\phi_{c}^{(n)}(w)$ are descendant fields of $\phi_c(w)$ of weight $h_c+n$, i.e. those built from linear combinations of fields of the form $(L_{-k_1}\ldots L_{-k_n}\phi)(w)$ for positive $k_i$ with $\sum_i k_i=n$.

The space of all descendant fields of a primary field $\phi_c(w)$ is the \emph{conformal family} $[\phi_c]$ of $\phi_c(w)$.  Under the state-field correspondence, the fields in a conformal family correspond precisely to vectors in the irreducible LWR built on the lowest weight vector $|\phi_c\rangle$.  In equation~\eqref{conformalfusion} it should be understood that more than one copy of each conformal family can appear in the sum on the right-hand side.

We record which conformal families appear in the fusion of $\phi_a(z)$ and $\phi_b(w)$ using the notation
\begin{align*}
  [\phi_a]\times[\phi_b]&\sim \sum_{c\in P}\; N_{ab}^c\;[\phi_c]
\end{align*}
where $N_{ab}^c\in\mZ$ counts the multiplicity of the family $[\phi_c]$ appearing on the right hand side.  The non-negative integers $N_{ab}^c$ are called the fusion rules of the theory.

In the $N=2$ case, the fusion between the super primary fields is \emph{a priori} again
\begin{align}\label{superconformalfusion}
  \phi_a(z)\phi_b(w)&=\sum_{c\in P}C_{ab}^c(z-w)^{h_c-h_a-h_b}\left[\phi_c(w)+\sum_{n>0}(z-w)^n\phi_c^{(n)}(w)\right]
\end{align}
where the $\phi_c(w)$ are $N=2$ descendant states (so, in particular, in the NS sector the sum over $n$ runs over positive half integers).  The fusion rules \emph{a priori} are
\begin{align*}
  [\phi_a]\times[\phi_b]&\sim \sum_{c\in P}N_{ab}^c\;[\phi_c].
\end{align*}
  We can view the OPE as a short-range expansion for fields inside a compatible system of $n$-point functions.  Then $J_0$ invariance of the $n$-point functions constrains the form of the OPE in equation~\eqref{superconformalfusion}.  It implies that the $U(1)$ charges of all the fields $\phi_c^{(n)}(w)$ must be equal.  This allows us to refine the fusion rules.  Descendants of $\phi_c(w)$ are of the form
\begin{align*}
    (L_{-n_1}\ldots L_{-n_{\alpha}}J_{-m_1}\ldots J_{-m_{\beta}}G^+_{-l_1}\ldots G^+_{-l_{\gamma}}G^-_{-k_1}\ldots G^-_{k_{\delta}}\phi_c)(w),
\end{align*}
which has $U(1)$ charge $Q_c+\frac{1}{2}(\gamma-\delta)$, where $Q_c$ is the $U(1)$ charge of $\phi_c(w)$.  We split the superconformal family $[\phi_c]$ into two subfamilies: $[c,+]$ containing those descendants with $\gamma-\delta$ even and $[c,-]$ containing those descendants with $\gamma-\delta$ odd.  We can then capture the interactions of the different even and odd superconformal `half-families' in the super fusion rules
\begin{align*}
  [a,\epsilon_a]\times[b,\epsilon_b]\sim \sum_{(c,\epsilon_c)\in P\times\{\pm\}} N_{(a,\epsilon_a)(b,\epsilon_b)}^{(c,\epsilon_c)}[c,\epsilon_c].
\end{align*}

We now specialise to the case of the $N=2$ minimal models.  Recall that the super-primary fields of the $N=2$ minimal models are labelled by those $(a,c)\in Q_k=\{0,\ldots,k\}\times\mZ_{2\overline{k}}$ that satisfy $|c-[a+c]|\leq a$.  According to the discussion in section~\ref{mi}, fields in $[(a,c),+]$ with $|c-[a+c]|\leq a$ correspond under the state-field correspondence precisely to states counted by the character $\chi_{ac}$, and fields in $[(a,c),-]$ to states counted by $\chi_{k-a,c+\overline{k}}$.  We will henceforth use the notation $[(a,c)]$ with $(a,c)\in Q_k$ to label the even and odd superconformal families for the $N=2$ minimal models.

The integers $N_{ac;a'c'}^{a''c''}$ calculated in lemma~\ref{fusionrules} are the natural candidates for the super fusion rules.  This result is confirmed by~\cite{Mussardo1988,Mussardo1988a} in the NS$\times$NS and R$\times$R sectors, both through the Coulomb gas formalism and through the explicit construction of the unitary $N=2$ minimal models via the parafermion-boson construction~\cite{Qiu1987}.  Furthermore, in section~\ref{fusionrulescheck}, we perform a non-trivial consistency check that all the possible modular invariants in Gannon's list are consistent with these fusion rules.

We can also read off the usual fusion between $N=2$ primary fields by simply forgetting the distinction between $[ac]$ and $[k-a,c+\overline{k}]$.  Then the fusion rules for the primary fields read
\begin{align*}
  \widehat{N}_{ac;a'c'}^{\alpha\gamma}&=N_{ac;a'c'}^{\alpha\gamma}+N_{ac;a'c'}^{k-\alpha,\gamma+\overline{k}}\\
  &=\left(N^{\widehat{\mathfrak{su}(2)}_k}\right)^{\alpha}_{aa'}\left(N^{\widehat{\mathfrak{u}(1)}_{\overline{k}}}\right)^{\gamma}_{cc'}+\left(N^{\widehat{\mathfrak{su}(2)}_k}\right)^{k-\alpha}_{aa'}\left(N^{\widehat{\mathfrak{u}(1)}_{\overline{k}}}\right)^{\gamma+\overline{k}}_{cc'}.
\end{align*}
It is precisely this quantity that Wakimoto calculates (using a Verlinde formula) in~\cite{Wakimoto1998}, and this agrees with the result of Adamovic~\cite{Adamovic2001}, which derives the fusion rules in the NS$\times$NS sector from the vertex operator point of view.  The author is not aware of corresponding vertex algebra results for the NS$\times$R, R$\times$NS and R$\times$R sectors.

In summary, the evidence presented in this section supports the following conjecture:
\begin{conjecture}\label{Verlindeformula}
  The fusion rules for the $N=2$ minimal models are given by
  \begin{align*}
    N_{ac;\,a'c'}^{\alpha\gamma}&=\sum_{(df)\in Q_k}\frac{S_{ac;df}S_{a'c';df}S_{\alpha\gamma;df}^*}{S_{00;df}}\\
    &=\left\{\begin{matrix}
    \left(N^{\widehat{\mathfrak{su}(2)}_k}\right)^{\alpha}_{aa'}\left(N^{\widehat{\mathfrak{u}(1)}_{\overline{k}}}\right)^{\gamma}_{cc'}, & \text{if  }[a+c][a'+c']=0\\
    \left(N^{\widehat{\mathfrak{su}(2)}_k}\right)^{k-\alpha}_{aa'}\left(N^{\widehat{\mathfrak{u}(1)}_{\overline{k}}}\right)^{\gamma+\overline{k}}_{cc'}, & \text{if  }[a+c][a'+c']=1
    \end{matrix}\right\}\\
    &\qquad\text{for }(ac),(a'c'),(\alpha\gamma)\in Q_k,
    \end{align*}
    where we label fields in the superconformal family of the super-primary $\phi_{ac}(z)$ with the same $U(1)$ charge as $\phi_{ac}(z)$ by $[ac]$, and fields whose $U(1)$ charge differs by a half integer by $[k-a,c+\overline{k}]$ for $|c-[a+c]|\leq a$.

    If we simply wish to label fields in the same superconformal family as $\phi_{ac}(z)$ by $[ac]$ then the fusion rules are
    \begin{align*}
      \widehat{N}_{ac;a'c'}^{\alpha\gamma}&=N_{ac;a'c'}^{\alpha\gamma}+N_{ac;a'c'}^{k-\alpha,\gamma+\overline{k}}\\
      &=\left(N^{\widehat{\mathfrak{su}(2)}_k}\right)^{\alpha}_{aa'}\left(N^{\widehat{\mathfrak{u}(1)}_{\overline{k}}}\right)^{\gamma}_{cc'}+\left(N^{\widehat{\mathfrak{su}(2)}_k}\right)^{k-\alpha}_{aa'}\left(N^{\widehat{\mathfrak{u}(1)}_{\overline{k}}}\right)^{\gamma+\overline{k}}_{cc'}
    \end{align*}
    for $(ac),(a'c'),(\alpha\gamma)\in\{(ln)\in Q_k\mid|n-[l+n]|\leq l\}$.
\end{conjecture}

\subsubsection{Simple Currents of the Minimal Models}\label{mmsc}
From the explicit formula for the fusion rules above one can read off that the simple currents of the minimal models at level $k$ are $\mathcal{J}=\{0,k\}\times\mZ_{2\overline{k}}$.  Each current acts naturally on the set of labels $Q_k$ of the $N=2$ minimal models: $\mathbf{j}$ maps $(a,c)$ to the label of the field which appears in the OPE of $\phi_{\mathbf{j}}$ and $\phi_{ac}$.  Thus, writing $J$ for the $\widehat{\mathfrak{su}(2)}_k$ current $J:a\mapsto k-a$, we have, for integer $l$,
\begin{align*}
  (J^l0,d)\cdot(a,c)&=(J^{l+(lk+d)(a+c)}a,c+d+(lk+d)(a+c)\overline{k}).
\end{align*}
The action of the currents on $Q_k$ defines an associative, commutative binary operation $\times$ on the set of currents by
\begin{align}\label{gplaw}
  \begin{split}
  (J^{l_1}0,d_1)\times(J^{l_2}0,d_2)&=(J^{l_1+l_2+(l_1k+d_1)(l_2k+d_2)}0,\\
  &\qquad d_1+d_2+(l_1k+d_1)(l_2k+d_2)\overline{k}).
  \end{split}
\end{align}
It is easy to check that $(0,0)$ is an identity element and that
\begin{align*}
  (J^l0,d)^{-1}&=(J^{l+lk+d}0,-d+(lk+d)\overline{k}).
\end{align*}
So the set of simple currents at level $k$ form a commutative group isomorphic to 
\begin{align*}
  \mathcal{J}&\cong
  \begin{cases}
    \mZ_{4\overline{k}} & \text{ if $k$ is odd}\\
    \mZ_2\times\mZ_{2\overline{k}} & \text{ if $k$ is even}.
  \end{cases}
\end{align*}
The simple currents are of great use because the $S$-matrix behaves well under the action of the currents on the weights.  In fact
\begin{align}
  S_{\mathbf{j}\cdot(ac);\,a'c'}&=\exp(2\pi iQ_{\mathbf{j}}(a',c'))S_{ac;\,a'c'}\label{simplecurrentaction}
\end{align}
where $Q_{(J^l0,d)}(a',c')=\frac{a'l}{2}+\frac{c'd}{2\overline{k}}-\frac{[kl+d][a'+c']}{4}$ and we have written $[b]\in\{0,1\}$ for the value of $b$ modulo 2, as before.  $Q_{\mathbf{j}}$ is called the \emph{monodromy charge} of the field $\phi_{ac}$ with respect to the current $\mathbf{j}$.  The monodromy charges satisfy
\begin{align*}
  Q_{\mathbf{j}}(a,c)&\equiv h_{\mathbf{j}}+h_{ac}-h_{\mathbf{j}\cdot(ac)} \mod \mZ,
\end{align*}
so $Q_{\mathbf{j}}(a,c)$ is also the monodromy of $\phi_{ac}$ with $\phi_\mathbf{j}$, as expected~\cite{Schellekens1989a}.

Note that in particular, \eqref{simplecurrentaction} applied to the simple current $(J0,\overline{k})$ gives
\begin{align}\label{simpleS}
  S_{Ja,c+\overline{k};\,a'c'}&=(-1)^{a'+c'}S_{ac;\,a'c'}.
\end{align}

\subsubsection{Simple Current Invariants}\label{sci}
It was observed in ~\cite{Kreuzer1994} that in all then-known cases, almost all the rational CFTs that can be constructed are the so-called \emph{simple current invariants}~\cite{GatoRivera1992}, leaving at worst a handful of ``exceptional'' models not of simple current type.  By simple current invariant we mean a CFT with partition function $Z=\sum_{l,l'}M_{l;l'}\chi_l\chi_{l'}^*$ where $\chi_l$ are the characters of the representations of the $\mathcal{W}$-algebra such that
\begin{align}\label{screlation}
  M_{l;l'}\neq 0\quad\Rightarrow\quad l'=\mathbf{j}\cdot l\text{ for some $\mathbf{j}\in\mathcal{J}$}
\end{align}
where $\mathcal{J}$ is the set of simple currents of the CFT.  This is a strong assumption indeed - see section 3 of~\cite{Gannon1995a} for a number of immediate consequences.

If we are interested in simple current invariants, then we are only concerned with those simple currents that can feature in equation~\eqref{screlation} for some modular invariant partition function.  $T$-invariance implies that we only need retain those currents whose conformal weight multiplied by their order is an integer.  To see this, let $\mathbf{j}$ be a current of order $n$ and suppose there exists an $l$ such that $M_{l;\,\mathbf{j}\cdot l}\neq 0$.  Then by $T$-invariance $h_l\equiv h_{\mathbf{j}l}\mod 1$, so $nh_{\mathbf{j}}\equiv nQ_{\mathbf{j}}(l)=Q_{\mathbf{j}^n}(l)=Q_{\text{id}}(l)=0\mod 1$.  Such currents form the \emph{effective centre}, $\mathcal{C}$~\cite{Kreuzer1994}.  In the case of the $N=2$ minimal models:
\begin{align*}
  \mathcal{C}_k=\begin{cases}
  \{(J^l0,d)\mid l+d\equiv0\!\!\!\mod2\}\cong\mZ_{2\overline{k}} & \text{if $k$ is odd,}\\
  \{0,k\}\times\{2d\mid d\in\mZ\}\cong\mZ_2\times\mZ_{\overline{k}} & \text{if $4|\overline{k}$,}\\
  \{0,k\}\times\mZ_{2\overline{k}}\cong\mZ_2\times\mZ_{2\overline{k}} & \text{if $4|k$,}
  \end{cases}
\end{align*}
which are groups under the group law inherited from~\eqref{gplaw}.

\subsection{Some Necessary Conditions}
At this point, we will prove two consistency checks of the minimal models, one pertaining to the fusion rules and one to the locality of the theory.

\subsubsection{Fusion Rules}\label{fusionrulescheck}
In section~\ref{verlindesection} we derived the chiral fusion rules of the minimal models.  The fusion rules enforce harsh restrictions on the OPE of a SCFT, so if a modular invariant $M$ really corresponds to the partition function of a minimal model, it must pass a consistency test imposed by the fusion rules.  This consistency test was performed in the case of $N=0$ minimal models by Gepner~\cite{Gepner1986b}.

Consider a possible theory with partition function corresponding to some modular invariant $M$.  If fields $\phi_{ac;a'c'}\in[ac]\otimes[a'c']$ and $\phi_{df;d'f'}\in[df]\otimes[d'f']$ are present then the fusion rules restrict the fusion between $\phi_{ac;a'c'}$ and $\phi_{df;d'f'}$ to lie in
\begin{align*}
  \sum_{\substack{(\alpha\gamma)\in Q_k\\(\alpha'\gamma')\in Q_k}}N_{ac;df}^{\alpha\gamma}N_{a'c';d'f'}^{\alpha'\gamma'}[\alpha\gamma]\otimes[\alpha'\gamma'].
\end{align*}
This expression is further constrained since only fields that show up in the partition function can be present\footnote{We remind the reader that the fusion rules give only an upper bound to the number fields produced under fusion of two fields -- it can easily happen that fewer fields appear than are allowed by the fusion rules.}.  If our theory is to be consistent, then we require that the fusion between any two fields is non-zero.  We confirm that the $N=2$ minimal models conform to this requirement in the following theorem:
\begin{theorem}\label{fusiontheorem}
  For any modular invariant $M$ in the list of Gannon (see section~\ref{minimalmodels}) we have
  \begin{align*}
    M_{ac;a'c'}\neq0,\;M_{df;d'f'}\neq 0&\implies N_{ac;df}^{\alpha\gamma}M_{\alpha\gamma;\alpha'\gamma'}N_{a'c';d'f'}^{\alpha'\gamma'}\neq0\\
    &\qquad\qquad\text{for some }(\alpha\gamma),(\alpha'\gamma')\in Q_k.
  \end{align*}
  This proves that the fusion rules do not preclude the existence of the $N=2$ minimal models.
\end{theorem}
\begin{proof}
  The fusion coefficients $N$ were given in lemma~\ref{fusionrules}.  One must work through the list of modular invariants~\eqref{M^0}-\eqref{E^{28}} checking the condition each time by hand.  The calculations are tedious and unenlightening, so they are not presented here.\qed
\end{proof}

\subsubsection{Locality}\label{locality}

In this section we will prove that theories corresponding to the modular invariants in Gannon's list have the expected locality properties.  In fact, locality follows from $T$-invariance of the theories, and we present the arguments generally for any SCFT.\footnote{I thank an anonymous referee for pointing out that my arguments apply more generally.}

We saw in equation~\eqref{Tcondition} that $T$-invariance of an SCFT is equivalent to the operator $e^{2\pi i(L_0-\overline{L}_0)}$ acting trivially on the bosonic part of the state space.  Since $L_0-\overline{L}_0$ generates rotations we have $\phi(e^{i\theta}z)=e^{i\theta(L_0-\overline{L}_0)}\phi(z)e^{-i\theta(L_0-\overline{L}_0)}$, and so all fields counted by the partition function are single-valued.

Furthermore, writing $M$ for the matrix of mulitplicities of a $T$-invariant partition function, we see that whenever $M_{\lambda,\lambda'},M_{\mu,\mu'}$ and $M_{\nu,\nu'}$ are all non-zero, we have $e^{2\pi i((h_{\lambda}-h_{\lambda'})+(h_{\mu}-h_{\mu'})-(h_{\nu}-h_{\nu'}))}=1$.  This proves that if $M$ permits the existence of a field $\phi_{\lambda;\lambda'}$ in the sector $(\lambda,\lambda')$, $\phi_{\mu;\mu'}$ in the sector $(\mu;\mu')$ and $\phi_{\nu;\nu;}$ in the sector $(\nu;\nu')$, and if the field $\phi_{\nu;\nu'}$ appears in the OPE of $\phi_{\lambda;\lambda'}$ with $\phi_{\mu;\mu'}$ then the OPE should be single-valued.

\section{Orbifold Construction of the \texorpdfstring{$N=2$}{N=2} Unitary Minimal Models}\label{orbifold}
In this section we establish the main result of this paper: the existence of a unitary $N=2$ minimal model for each possible partition function.  The statement of the result is given formally in section~\ref{existence}.  The proof rest upon the existence of orbifoldings between the space-time supersymmetric \cADE models and the less familiar models given in Gannon's list (see section~\ref{minimalmodels}).  The main step of the proof is to prove the following theorem:
\begin{theorem}\label{theorem1}
  \mbox{}
  \begin{itemize}
  \item  Every non-exceptional partition function of a unitary $N=2$ minimal model at level $k$ can be obtained by orbifoldings of the diagonal partition function at level $k$.
  \item  Every exceptional partition function of a unitary $N=2$ minimal model with level $k=10$, $16$ or $28$ can be obtained by orbifoldings of the $\mathcal{E}_6\otimes I_{24}$, $\mathcal{E}_7\otimes I_{36}$ or $\mathcal{E}_8\otimes I_{60}$ partition functions, respectively, where $\mathcal{E}_{6,7,8}$ are the $\widehat{\mathfrak{su}(2)}_k$ exceptional modular invariants, and $I_{2\overline{k}}$ is the $\widehat{\mathfrak{u}(1)}_{\overline{k}}$ diagonal invariant.
  \end{itemize}
\end{theorem}

We will prove this theorem, by explicitly constructing the necessary orbifoldings, in section~\ref{pf1}.  We must first explain what we mean by \emph{orbifolding}.

\subsection{Orbifolding}
We first describe the orbifolding procedure\footnote{We note here that some authors (e.g.~\cite{ffrs09}) use the term `orbifolding' to mean restricting the underlying vertex operator algebra to the fixed point of a group action to arrive at a new vertex operator algebra.  In this paper, we use the notion of orbifolding found in e.g.~\cite{Ginsparg1988}, and only deal with orbifoldings that do not take us away from the category of representations of the super Virasoro algebra, as explained below.} in the case of a rational (bosonic) CFT.  Let $\mH$ be the underlying pre-Hilbert space of a CFT $\mathcal{C}$ and let $\rho:G\to\text{End}(\mH)$ be an action of a finite group $G$ on $\mH$ such that
\begin{enumerate}
\item $\mH$ is simultaneously diagonalisable with respect to $L_0,\overline{L}_0$ and $\rho(g)$ for every $g\in G$, where $L_0,\overline{L}_0$ are viewed as linear operators on $\mH$;
\item $\rho(g)$ commutes with $L_n$ and $\overline{L}_n$ for every $n$, where $L_n,\overline{L}_n$ are viewed as linear operators on $\mH$.
\end{enumerate}
In this paper we will only be interested in the simple case where $\rho(g)$ acts by multiplication by a root of unity on each irreducible representation of the extended symmetry algebra for all $g\in G$.  This will always be the case when for example the lowest weight space is 1-dimensional.  Decompose $\mH=\bigoplus_{a\,\in P_l,b\,\in P_r}\mH_a\otimes\overline{\mH}_b$, where $P_l,P_r$ are sets of labels of (not necessarily distinct) irreducible representations of the symmetry algebra.  $\rho(g)$ then acts by multiplication by the root of unity $\xi_{a,b}(g)$ on the irreducible representation $\mH_a\otimes\overline{\mH}_b$.  It follows that the action of $G$ on the states of $\mH$ is entirely described by its action on the characters $\rho(g)(\chi_{a}\chi_{b}^*)=\xi_{a,b}(g)\chi_{a}\chi_{b}^*$.  For notational simplicity we shall now simply write $g$ in place of $\rho(g)$.

We want to construct a $G$-invariant CFT from $\mathcal{C}$, the \emph{$G$-orbifold} of $\mathcal{C}$, denoted $\mathcal{C}/G$.  We will restrict our attention to an abelian group $G$ for ease of notation, but one can generalise to non-abelian groups with a little care (see e.g.~\cite{Ginsparg1988}).

We begin by projecting onto the $G$-invariant states of $\mathcal{C}$:
\begin{align*}
  \mH^{\text{inv}}:=\mathcal{P}\cdot\mH
\end{align*}
where the projector $\mathcal{P}$ is given by $\frac{1}{|G|}\sum_{g\in G}g\cdot$.  We use a notational shorthand
\begin{align*}
  \orbox{g}{1}&:=\text{Tr}_{\mH}(gq^{L_0-\frac{\mathbf{\bar{c}}}{24}}\overline{q}^{\overline{L}_0-\frac{\mathbf{\bar{c}}}{24}})
\end{align*}
for the trace with $g$ inserted, which makes sense because of condition 1 above.  This allows us to write the partition function of the $G$-invariant sector as
\begin{align*}
  Z^{\text{inv}}(\tau)&=\text{Tr}_{\mH}(\mathcal{P}q^{L_0-\frac{\mathbf{\bar{c}}}{24}}\overline{q}^{\overline{L}_0-\frac{\mathbf{\bar{c}}}{24}})=\frac{1}{|G|}\sum_{g\in G}\orbox{g}{1}.
\end{align*}
Unless $G$ is trivial, $Z^{\text{inv}}(\tau)$ will not be modular invariant.  In order to restore modular invariance we need to add in extra $G$-invariant states, the so called \emph{twisted states}.

The problem of constructing the twisted states is difficult in general, but we will only be interested in the case of the unitary $N=2$ minimal models.  In this case we can construct the twisted sector out of known representations, using the following arguments: by condition 2, the $L_n,\overline{L}_n$ modes commute with the $G$-action and so the central charge $\mathbf{\bar{c}}$ is left invariant, and since the action of $\text{SL}(2,\mZ)$ leaves $\mathbf{\bar{c}}$ invariant the twisted sector should also be composed of irreducible representations at central charge $\mathbf{\bar{c}}$.  But in the situation of interest to us, the collection of irreducible representations are explicitly known for fixed $\mathbf{\bar{c}}$.  Thus the twisted sector can be constructed from these known representations.  It is therefore sufficient to find the partition function of the twisted sector using standard tricks below.

We now return to the construction of the partition function of the twisted sector.  For each $h\in G$ we denote by $\mH_h$ the sector of states `twisted by $h$' in the space direction; in the language of fields we make a cut from $0$ to $\tau$ along the world-sheet torus $T=\mC/(\mZ\oplus\tau\mZ)$ and require that a field crossing the cut is acted on by $h$:
\begin{align*}
  \phi(z+1)&=h\phi(z).
\end{align*}
Since we want to keep only $G$-invariant states, we project the partition function of $\mH_h$ with $\mathcal{P}$:
\begin{align*}
  \text{Tr}_{\mH_h}(\mathcal{P}q^{L_0-\frac{\mathbf{\bar{c}}}{24}}\overline{q}^{\overline{L}_0-\frac{\mathbf{\bar{c}}}{24}})=\frac{1}{|G|}\sum_{g\in G}\orbox{g}{h},
\end{align*}
where we have introduced the notational shorthand
\begin{align*}
  \orbox{g}{h}:=\text{Tr}_{\mH_h}(gq^{L_0-\frac{\mathbf{\bar{c}}}{24}}\overline{q}^{\overline{L}_0-\frac{\mathbf{\bar{c}}}{24}}).
\end{align*}
Then the partition function of the orbifold theory is the sum of the contributions from each of the twisted sectors:
\begin{align}
  Z^{\text{orb}}&=\frac{1}{|G|}\sum_{g,h\in G}\orbox{g}{h}.\label{orb}
\end{align}
We interpret the box $\orbox{g}{h}$ as counting states whose fields live on the world-sheet torus with a cut along each cycle, such that cycling around once in the space-direction yields a factor of $h$ and cycling around once in the time-direction yields a factor of $g$:
\begin{align*}
  \phi(z+1)&=h\phi(z),\\
  \phi(z+\tau)&=g\phi(z).  
\end{align*}
Then we find that the $S$ and $T$-transformations act to permute the `boundary conditions' in the following way:
\begin{align*}
  S\left(\orbox{g}{h}\right)&=\orbox{h^{-1}}{g}\;,\\
  T\left(\orbox{g}{h}\right)&=\orbox{gh}{h}\;,
\end{align*}
thus ensuring modular invariance of the orbifold partition function.

Actually, we have slightly greater freedom in piecing together the $\text{SL}(2,\mZ)$ orbits than we have shown in equation~\eqref{orb}, since we can introduce phases between the different orbits and still arrive at something modular invariant.  The freedom we have in choosing these phases is called \emph{discrete torsion} and is classified by the second group cohomology $H^2(G,U(1))$~\cite{Vafa1986}\footnote{As Vafa points out in~\cite{Vafa1986}, the relative phase between orbits in a 1-loop modular invariant is actually given by a function $\epsilon(g,h):=w(g,h)w(h,g)^{-1}$ for cocycles $w\in H^2(G,\mC^*)$.}.  In this paper we will need to consider only the cases $G=\mZ_k$ (for which $H^2(\mZ_k,U(1))\cong\mZ_1$ contains only a single class) and $G=\mZ_2\times\mZ_{2k}$ (with discrete torsion $H(\mZ_2\times\mZ_{2k},U(1))\cong\mZ_2$ consists of two distinct classes).

This completes the construction for bosonic CFTs.  In order to extend the prescription to the SCFT case, we just replace the space of states $\mH$ with the bosonic states, and add the $z$-dependence (via $y^{J_0}$) into the traces in the obvious manner.

We note here that although modular invariance is guaranteed, it may well be that the resulting partition function is not valid, in the sense that it may not have non-negative integer coefficients when viewed as a state-counting formal power series in $q,\overline{q}$, or in the sense that it may not correspond to a consistent full CFT.  It will be evident that the orbifolds in this section pass the first of these tests.  That they do not fail the second test is the content of the following assumption:
\begin{assumption}\label{assumption1}
  Orbifolds of a physical theory are again physical.
\end{assumption}
By this we mean that an orbifolding of a fully-fledged SCFT that satisfies the two conditions given at the beginning of this section, will give rise to another fully-fledged SCFT.  Up to now, the orbifoldings we have constructed have been given entirely in terms of the partition function.  In order to have a chance of getting an orbifold SCFT we must impose the \emph{level-matching conditions}~\cite{Vafa1986,Dixon1986}; that is, we must check that the spin $h-\overline{h}$ of the fields in the orbifold theory remain at worst half-integral and also that we do not destroy semi-locality of the fields.  We will explicitly see in section~\ref{pf1} that in all the orbifolds we consider, we obtain another partition function from Gannon's list.  But we know from equation~\eqref{Tcondition} and section~\ref{locality} that all states counted by the partition functions have integral spin and are mutually local.  Since the spins of states in the full Hilbert space differ at worst by a half-integer from spins of these states, we see that all states have integral or half-integral spin and are at worst mutually semi-local .

\subsection{Existence of the \texorpdfstring{$N=2$}{N=2} Minimal Models}\label{existence}
We require one final assumption before we can state and prove the existence of the $N=2$ minimal models corresponding to each of Gannon's partition functions.
\begin{assumption}\label{assumption2}
  The space-time supersymmetric $\mathcal{A},\mathcal{D}$ and $\mathcal{E}$ models are fully-fledged, physical SCFTs.
\end{assumption}
By this we mean we adopt the (widely believed) assumption that the $\mathcal{A},\mathcal{D}$ and $\mathcal{E}$ models are genuine SCFTs that admit a consistent system of $n$-point correlators on Riemann surfaces of all genera.  This is at least partially known to be true; for example, the genus zero OPE coefficients of the $\mathcal{A}$ model were calculated in~\cite{Mussardo1988a} using the relation between the parafermion fields with those of the $\mathfrak{su}(2)$ WZW models~\cite{Zamolodchikov1986a}, and the OPE coefficients of the exceptional models should in principal be calculable using the free field construction of e.g.~\cite{Feigin1998}.  Furthermore, it was shown in~\cite{MS1988,MS1989b,Sonoda1988} that a (bosonic) rational theory admits a consistent system of $n$-point correlators on Riemann surfaces of all genera if it admits a consistent system of $4$-point correlators on the sphere and $1$-point correlators on the torus.

Alternatively, in the framework of topological quantum field theories and modular tensor categories, it was shown in~\cite{frs02} that in the categories of modules over a vertex operator algebra, the `Cardy case' is always realised.  For the $N=2$ minimal models, this is nothing other than the mirror partner of the $\mathcal{A}$-model, which is physically equivalent to the $\mathcal{A}$-model itself.

\begin{theorem}
  Given assumptions~\ref{assumption1} and~\ref{assumption2}, there corresponds to each of the candidate partition functions given in Gannon's list (see section~\ref{minimalmodels}) a fully-fledged superconformal field theory.
\end{theorem}
\begin{proof}
  Theorem~\ref{theorem1} shows that every partition function is obtained from one of a handful of possible partition functions by a chain of orbifoldings by cyclic groups.  Since orbifoldings by solvable groups can be inverted (see e.g.~\cite{Ginsparg1988}) it follows that we can obtain by a chain of orbifoldings any given partition function from the $\mathcal{A}$ model (if it is a simple current invariant), or from the $\mathcal{E}_6,\mathcal{E}_7,\mathcal{E}_{8}$ model (if it is an exceptional invariant with $k=10,16,28$ respectively).

  The existence of the $N=2$ minimal models then follows immediately from assumptions~\ref{assumption1} and~\ref{assumption2}.\qed
\end{proof}

\subsection{Proof of Theorem~\ref{theorem1}}\label{pf1}
  The proof is constructive: given any modular invariant $M$ at level $k$ in Gannon's list, we construct a chain of orbifoldings (by cyclic groups) mapping $M$ to either the $\cal{A}$ or the $\cal{E}$ space-time supersymmetric minimal model.

The proof will be broken down into several sections.  In sections~\ref{O^1} and~\ref{O^2} we will introduce some simple $\mZ_2$ orbifolds which realise certain global symmetries discussed briefly in section~\ref{symm}.  In section~\ref{O^3} we generalise a well-known $\mZ_2$ orbifold from the $\widehat{\mathfrak{su}(2)}_k$ models to the minimal models, and observe that we can construct an orbifolding between the minimal ``families'' listed in section~\ref{minimalmodels}.

In sections~\ref{between}-\ref{O^5andO^6} we state and prove a proposition that every modular invariant $M$ can be mapped into either $\widetilde{M}^{0},\widetilde{M}^{4,2},\widetilde{M}^{2,0},\widetilde{E}^{10}_1,\widetilde{E}^{16}_2$ or $\widetilde{E}^{28}$ depending on the level $k$ and whether $M$ is exceptional or not.

We then attempt to control the parameter $v$ -- we find an orbifolding to map any given modular invariant in one of the above families to the modular invariant with the lowest possible value of $v$.  This is sections~\ref{vcontrol} to~\ref{vcontrolend}.

Lastly, in sections~\ref{zcontrol}-~\ref{zcontrolend} we try to control the parameter $z$.  We summarise these results in section~\ref{summary}, finally completing the proof.

In order to cut out pages of technical proofs, we will in general just write down the general `box' $\orbox{g}{h}$ for $g,h\in G$ for an orbifolding, observe that it gives the expected result when $h=0$, and state the resulting orbifold partition function.  The behaviour under modular transformations will be shown to be correct only for the first simple examples, since the proof is similar in the other cases.  The reader who wants more detailed proofs should consult~\cite{Gray2009}.

\subsubsection{The Orbifoldings \texorpdfstring{$\mathcal{O}^1_L$}{O1L},\texorpdfstring{$\mathcal{O}^1_R$}{O1R}}\label{O^1}
Let $Z\equiv Z(\tau,z)$ be a modular invariant from the list in section~\ref{minimalmodels}.  We write
\begin{align*}
  Z=\orbox{1}{1}&=\sum_{\substack{(ac)\in Q_k\\(a'c')\in Q_k}}M_{ac;\,a'c'}\chi_{ac}\chi_{a'c'}^*
\end{align*}
and let $\mZ_2=\langle g\rangle$ act on the states via
\begin{align}
  g\cdot\chi_{ac}\chi_{a'c'}^*&=(-1)^{a+c}\chi_{ac}\chi_{a'c'}^*.\label{Z2action}
\end{align}
Since the parity of $a+c$ determines whether the states counted by $\chi_{ac}$ are in the NS or R sectors, we see that this action leaves the NS sector invariant.  The general box for $m,n\in\{0,1\}$ is given by
\begin{align*}
  \orbox{g^m}{g^n}&=\sum_{\substack{(ac)\in Q_k\\(a'c')\in Q_k}}M_{\mathbf{j}^n(ac);\,a'c'}(-1)^{(a+c+n)m}\chi_{ac}\chi_{a'c'}^*,
\end{align*}
where from now on $\mathbf{j}(ac)=(k-a,c+\overline{k})$.  This is clearly correct when $n=0$, and since there is no discrete torsion, it remains to check that the general box transforms correctly under the $S$- and $T$-transformations.

For the $T$-transformation we find
\begin{align*}
  T\cdot\orbox{g^m}{g^n}&=\sum_{\substack{(ac)\in Q_k\\(a'c')\in Q_k}}M_{\mathbf{j}^n(ac);\,a'c'}(-1)^{(a+c+n)m}e^{2\pi i(h_{a,c}-h_{a',c'})}\chi_{ac}\chi_{a'c'}^*\\
  &=\sum_{\substack{(ac)\in Q_k\\(a'c')\in Q_k}}M_{\mathbf{j}^n(ac);\,a'c'}(-1)^{(a+c+n)m}\\
  &\qquad \times(-1)^{(a+c+1)n}e^{2\pi i(h_{\mathbf{j}^n(a,c)}-h_{a',c'})}\chi_{ac}\chi_{a'c'}^*\\
  &=\sum_{\substack{(ac)\in Q_k\\(a'c')\in Q_k}}M_{\mathbf{j}^n(ac);\,a'c'}(-1)^{(a+c+n)(m+n)}\chi_{ac}\chi_{a'c'}^*\\
  &=\orbox{g^{m+n}}{g^n}\;,
\end{align*}
where we used equation~\eqref{T}, then equation~\eqref{hq} and then equation~\eqref{Tcondition}.

For the $S$-matrix, we can simplify the calculation enormously if we use our knowledge of its behaviour under the action of simple currents (see section~\ref{simplecurrents}).  We find that
\begin{align*}
  S\cdot\orbox{g^m}{g^n}&=\sum_{\substack{(ac)\in Q_k\\(a'c')\in Q_k}}\sum_{\substack{(rs)\in Q_k\\(tu)\in Q_k}}S_{rs;\,ac}M_{\mathbf{j}^n(ac);\,a'c'}S_{a'c';\,tu}^*(-1)^{(a+c+n)m}\chi_{rs}\chi_{tu}^*\\
  &=\sum_{\substack{(ac)\in Q_k\\(a'c')\in Q_k}}\sum_{\substack{(rs)\in Q_k\\(tu)\in Q_k}}S_{rs;\,\mathbf{j}^n(ac)}M_{ac;\,a'c'}S_{a'c';\,tu}^*(-1)^{(a+c+n)m}\chi_{rs}\chi_{tu}^*\\
  &=\sum_{\substack{(ac)\in Q_k\\(a'c')\in Q_k}}\sum_{\substack{(rs)\in Q_k\\(tu)\in Q_k}}\!\!\!\!S_{rs;\,ac}M_{ac;\,a'c'}S_{a'c';\,tu}^*\\
  &\qquad\qquad\qquad\qquad\qquad\qquad\times(-1)^{(a+c+n)m+(r+s)n}\chi_{rs}\chi_{tu}^*\\
  &=\sum_{\substack{(ac)\in Q_k\\(a'c')\in Q_k}}\sum_{\substack{(rs)\in Q_k\\(tu)\in Q_k}}S_{\mathbf{j}^m(rs);\,ac}M_{ac;\,a'c'}S_{a'c';\,tu}^*(-1)^{(r+s+m)n}\chi_{rs}\chi_{tu}^*\\
  &=\sum_{\substack{(rs)\in Q_k\\(tu)\in Q_k}}M_{\mathbf{j}^m(rs);\,tu}(-1)^{(r+s+m)n}\chi_{rs}\chi_{tu}^*\\
  &=\orbox{g^{-n}}{g^m}\;,
\end{align*}
where in the third and fourth lines we used the nice behaviour of the $S$-matrix under the action of simple current $\mathbf{j}=(J0,\overline{k})$ given in equation~\eqref{simpleS}, and in the fifth line we used equation~\eqref{Scommute}.  Thus the boxes transform correctly under the action of $\text{SL}(2,\mZ)$.  Summing over all four boxes and multiplying by $\frac{1}{2}$ produces
\begin{align*}
  Z^{\text{orb}}&=\sum_{\substack{(ac)\in Q_k\\(a'c')\in Q_k\\a+c\text{ even}}}M_{ac;\,a'c'}\chi_{ac}\chi_{a'c'}^*+\sum_{\substack{(ac)\in Q_k\\(a'c')\in Q_k\\a+c\text{ odd}}}M_{\mathbf{j}(ac);\,a'c'}\chi_{ac}\chi_{a'c'}^*.
\end{align*}
This orbifolding defines an involution on the set of modular invariants.  We will refer to this orbifolding as $\mathcal{O}^1_L$ (where the L stands for left).  Since $S$ and $T$ are symmetric, it is clear that we could equally well have let $\mZ_2$ act on the right-hand representations, $g\cdot\chi_{ac}\chi_{a'c'}^*=(-1)^{a'+c'}\chi_{ac}\chi_{a'c'}^*$.  The result would be
\begin{align*}
  Z^{\text{orb}}&=\sum_{\substack{(ac)\in Q_k\\(a'c')\in Q_k\\a'+c'\text{ even}}}M_{ac;\,a'c'}\chi_{ac}\chi_{a'c'}^*+\sum_{\substack{(ac)\in Q_k\\(a'c')\in Q_k\\a'+c'\text{ odd}}}M_{ac;\,\mathbf{j}(a'c')}\chi_{ac}\chi_{a'c'}^*.
\end{align*}
We will refer to this orbifolding as $\mathcal{O}^1_R$.  

The reason we have done this relatively simple example in such great detail is that the procedure for checking $\text{SL}(2,\mZ)$-invariance for all other orbifoldings in this paper is very similar: one directly checks $T$-invariance with the help of equation~\eqref{Tcondition} and then uses the simple current action on the $S$-matrix to check $S$-invariance.  For an orbifolding with a cyclic group $G$, there is no discrete torsion, so the unique orbifold partition function is given by $\frac{1}{|G|}$ multiplied by the sum of the boxes.

\subsubsection{The Orbifoldings \texorpdfstring{$\mathcal{O}^2_L$}{O2L},\texorpdfstring{$\mathcal{O}^2_R$}{O2R}}\label{O^2}
Again we start with a minimal model with partition function
\begin{align*}
  Z=\orbox{1}{1}&=\sum_{\substack{(ac)\in Q_k\\(a'c')\in Q_k}}M_{ac;\,a'c'}\chi_{ac}\chi_{a'c'}^*
\end{align*}
and define a group action by $g\cdot\chi_{ac}\chi_{a'c'}^*=e^{\frac{2\pi ic}{\overline{k}}}\chi_{ac}\chi_{a'c'}^*$.  This defines a $\mZ_{\overline{k}}\,$-action.  We claim that the general box for $m,n\in\{0,\ldots,\overline{k}-1\}$ is given by
\begin{align*}
  \orbox{g^m}{g^n}&=\sum_{\substack{(ac)\in Q_k\\(a'c')\in Q_k}}M_{ac;\,a'c'}e^{\frac{2\pi im(c-n)}{\overline{k}}}\chi_{a,c-2n}\chi_{a'c'}^*.
\end{align*}
One easily checks that this is correct when $n=0$.  One checks that it transforms correctly under the $S$ and $T$ transformations just as in the previous case\footnote{The step-by-step calculations for this and some other orbifolds can be found in the author's thesis~\cite{Gray2009}.}: for the $T$ transformation, use equations~\eqref{T}, \eqref{hq} and then~\eqref{Tcondition} to show that
\begin{align*}
  T\cdot\orbox{g^m}{g^n}&=\orbox{g^{m+n}}{g^n}
\end{align*}
and for the $S$-transformation use equations~\eqref{simplecurrentaction} and~\eqref{Scommute} to show that
\begin{align*}
  S\cdot\orbox{g^m}{g^n}&=\orbox{g^{-n}}{g^m}.
\end{align*}
Thus the boxes span a representation of $\text{SL}(2,\mZ)$.  To find the resulting orbifold we calculate
\begin{align*}
  Z^{\text{orb}}&=\frac{1}{\overline{k}}\sum_{n,m=0,\ldots,\overline{k}-1}\orbox{g^m}{g^n}\\
  &=\sum_{\substack{(ac)\in Q_k\\(a'c')\in Q_k}}M_{a,-c;\,a'c'}\chi_{ac}\chi_{a'c'}^*.
\end{align*}
This orbifolding is well-defined on all minimal modular invariants.  We will refer to it by $\mathcal{O}^2_L$.  The group $\mZ_{\overline{k}}$ could equally as well have acted upon the right-hand representations.  In that case we would obtain
\begin{align*}
  Z^{\text{orb}}&=\sum_{\substack{(ac)\in Q_k\\(a'c')\in Q_k}}M_{ac;\,a',-c'}\chi_{ac}\chi_{a'c'}^*.
\end{align*}
We will refer to this orbifolding as $\mathcal{O}^2_R$.  Clearly these orbifoldings give the same result if the initial modular invariant is symmetric.

\subsubsection{Symmetries generated by \texorpdfstring{$\mathcal{O}^1_{L,R}$}{O1LR} and \texorpdfstring{$\mathcal{O}^2_{L,R}$}{O2LR}}\label{symm}
Note that these orbifoldings are self-inverse, they are mutually commuting, and the effect of concatenating $\mathcal{O}^1_L\mathcal{O}^2_L$ or $\mathcal{O}^1_R\mathcal{O}^2_R$ is to perform the mirror symmetry transformation on the left- or right-chiral half of the theory, respectively:
\begin{align*}
  \mathcal{O}^1_L\mathcal{O}^2_L&:M_{ac;\,a'c'}\mapsto M_{\mathbf{j}^{a+c}(a,-c);\,a'c'},\\
  \mathcal{O}^1_R\mathcal{O}^2_R&:M_{ac;\,a'c'}\mapsto M_{ac;\,\mathbf{j}^{a'+c'}(a',-c')},
\end{align*}
where left- or right-handed mirror symmetry is defined by performing charge conjugation on the left- or right-handed representations, respectively.  In terms of the partition functions, it is realised by multiplication of the modular invariant $M$ by the permutation matrix $S^2$ on the left or right respectively.  Using equation~\eqref{hq}, one checks that making the transformation $(a,c)\rightarrow\mathbf{j}^{a+c}(a,-c)$ has the effect of sending
\begin{align*}
  (h_{ac},Q_{ac})\rightarrow(h_{ac},-Q_{ac})\mod1
\end{align*}
as expected.

Performing charge conjugation on both sides simultaneously amounts to performing all 4 orbifoldings $\mathcal{O}^1_L\mathcal{O}^2_L\mathcal{O}^1_R\mathcal{O}^2_R$ in succession.  Since $S^4=\text{Id}$ and modular invariants commute with $S$, this has no overall effect on the partition function.  As discussed in section~\ref{k=1}, we consider two charge conjugate models (i.e. related by simultaneous charge conjugation on both chiral halves of the theory) to be equivalent; indeed they have the same partition function.  We will however not consider the mirror symmetry pairs to be equivalent in this paper, since they generally have distinct partition functions.

The results of applying $\mathcal{O}^1_{L,R}$ and $\mathcal{O}^2_{L,R}$ to the minimal partition functions listed in section~\ref{minimalmodels} are given in table~\ref{O^2table}.\footnote{The parameter $z$ is defined modulo some number $\alpha$ in each case.  $-z$ is to be understood as $-z\mod\alpha$.}

\begin{table}
  \caption{Action of $\mathcal{O}^1_{L,R},\mathcal{O}^2_{L,R}$ on minimal partition functions}
  \smallskip
  \centering
  \label{O^2table}
  \resizebox{!}{3cm}{
    \begin{tabular}[c]{ccccccc}
      \toprule
      \multicolumn{2}{c}{} & Id & $\mathcal{O}^1_L$ & $\mathcal{O}^1_R$ & $\mathcal{O}^2_L$ & $\mathcal{O}^2_R$\\ \midrule
      $k$ odd & $\widetilde{M}^0$ & $(v,z,n)$ & $(v,z,n+1)$ & $(v,z,n+1)$ & $(v,-z,n)$ & $(v,-z,n)$\\ \midrule
      \multirow{3}{*}{4 divides $\overline{k}$} & $\widetilde{M}^{2,0}$ & $(v,z,n)$ & $(v,z,n+1)$ & $(v,z,n+1)$ & $(v,-z,n)$ & $(v,-z,n)$\\\cmidrule{2-7}
      & $\widetilde{M}^{2,1}$ & $(v,z,n)$ & $(v,z,n+1)$ & $(v,z,n+1)$ & $(v,-z,n+1)$ & $(v,-z,n+1)$\\\cmidrule{2-7}
      & $\widetilde{M}^{2,2}$ & $(v,z,n,m)$ & $(v,z,n+1,m+1)$ & $(v,z,n+1,m+1)$ & $(v,-z,n,m)$ & $(v,-z,n,m)$\\ \midrule
      \multirow{4}{*}{4 divides $k$} & $\widetilde{M}^{4,0}$ & $(v,z,n,m)$ & $(v,z,n,m+1)$ & $(v,z,n+1,m)$ & $(v,-z,n+1,m)$  & $(v,-z,n,m+1)$\\\cmidrule{2-7}
      & $\widetilde{M}^{4,1}$ & $(v,z,x,y)$ & $(v,z,x+2,y)$ & $(v,z,x,y+2)$ & $(v,-z,x+2,y)$ & $(v,-z,x,y+2)$\\\cmidrule{2-7}
      & $\widetilde{M}^{4,2}$ & $(v,z,x)$ & $(v,z,x+2)$ & $(v,z,x+2)$ & $(v,-z,x+2)$ & $(v,-z,x+2)$\\\cmidrule{2-7}
      & $\widetilde{M}^{4,3}$ & $(v,z,n)$ & $(v,z,n+1)$ & $(v,z,n+1)$ & $(v,-z,n)$ & $(v,-z,n)$\\ \midrule
      \multirow{2}{*}{$k=10$} & $\widetilde{E}^{10}_1$ & $(6,z)$ & $(6,z)$ & $(6,z)$ & $(6,-z)$ & $(6,-z)$\\\cmidrule{2-7}
      & $\widetilde{E}^{10}_2$ & $(12,z,0,m)$ & $(12,z,0,m+1)$ & $(12,z,0,m+1)$ & $(12,-z,0,m)$ & $(12,-z,0,m)$\\ \midrule
      \multirow{2}{*}{$k=16$} & $\widetilde{E}^{16}_1$ & $(v,z,x,y)$ & $(v,z,x+2,y)$ & $(v,z,x,y+2)$ & $(v,-z,x+2,y)$ & $(v,-z,x,y+2)$\\\cmidrule{2-7}
      & $\widetilde{E}^{16}_2$ & $(v,z,x)$ & $(v,z,x+2)$ & $(v,z,x+2)$ & $(v,-z,x+2)$ & $(v,-z,x+2)$\\ \midrule
      $k=28$ & $\widetilde{E}^{28}$ & $(15,z,x)$ & $(15,z,x+2)$ & $(15,z,x+2)$ & $(15,-z,x+2)$ & $(15,-z,x+2)$\\ \bottomrule
    \end{tabular}
  }
\end{table}

\subsubsection{The generalised \texorpdfstring{$\mathcal{A}_k\leftrightarrow\mathcal{D}_k$}{AkDk} Orbifolding}\label{O^3}
The family $\widetilde{M}^{2,2}$ exists for any $k$ with $4|\overline{k}$.  Given such a $k$, we can always choose $v=\overline{k}$ and $z=1$.  Then, from equation~\eqref{M^{2,2}}, we obtain a modular invariant $M$ with $M_{ac;\,a'c'}=\delta(a'=J^{an}a)\delta(c'=c)$.  Thus
\begin{align*}
  M=\begin{cases}
  \mathcal{A}_k\otimes\mathcal{I}_{2\overline{k}} & \text{if $n=0$}\\
  \mathcal{D}_k\otimes\mathcal{I}_{2\overline{k}} & \text{if $n=1$},
  \end{cases}
\end{align*}
where the $\mathcal{A}$ and $\mathcal{D}$ are the partition functions of the $\widehat{\mathfrak{su}(2)}_k$ models of the same name encountered in~\cite{Cappelli1987} and $\mathcal{I}_{2\overline{k}}$ is the diagonal $\widehat{\mathfrak{u}(1)}_{\overline{k}}$ invariant.  Similarly, when $4$ divides $k$, the modular invariant $\widetilde{M}^{4,3}$ with parameters $v=\frac{\overline{k}}{2}$, $z=1$ and $n=0$ yields $\mathcal{A}_k\otimes\mathcal{I}_{2\overline{k}}$ and the modular invariant $\widetilde{M}^{4,2}$ with $v=\frac{\overline{k}}{2}$, $z=1$ and $x=1$ yields $\mathcal{D}_k\otimes\mathcal{I}_{2\overline{k}}$, where again the $\mathcal{A}$ and $\mathcal{D}$ are the partition functions of the $\widehat{\mathfrak{su}(2)}_k$ classification.
  Inspired by the well-known $\mZ_2$ orbifolding between the $\mathcal{A}$- and $\mathcal{D}$-models (see e.g.~\cite{DiFrancesco1997}), we define a $\mZ_2$ action on the states of an arbitrary modular invariant with even $k$ by
\begin{align*}
  g\cdot\chi_{ac}\chi_{a'c'}^*&:=(-1)^a\chi_{ac}\chi_{a'c'}^*.
\end{align*}
Then we find
\begin{align*}
  \orbox{g^m}{g^n}&=\sum_{\substack{(ac)\in Q_k\\(a'c')\in Q_k}}M_{ac';\,a'c'}(-1)^{(a+\frac{nk}{2})m}\chi_{J^na,c}\chi_{a'c'}^*.
\end{align*}
Thus
\begin{align*}
  Z^{\text{inv}}&=\sum_{\substack{(ac)\in Q_k\\a\equiv 0\!\!\!\!\!\mod2}}\sum_{(a'c')\in Q_k}M_{ac;\,a'c'}\chi_{ac}\chi_{a'c'}^*,\\
  Z^{\text{twist}}&=\sum_{\substack{(ac)\in Q_k\\a\equiv \frac{k}{2}\!\!\!\!\!\mod2}}\sum_{(a'c')\in Q_k}M_{Ja,c;\,a'c'}\chi_{ac}\chi_{a'c'}^*,\\
  Z^{\text{orb}}&=\begin{cases}
    \sum_{\substack{(ac)\in Q_k\\(a'c')\in Q_k}}M_{J^aa,c';\,a'c'}\chi_{ac}\chi_{a'c'}^*,\quad\text{if }4|\overline{k};\\
    \sum_{\substack{(ac)\in Q_k\\(a'c')\in Q_k\\a\text{ even}}}(M_{ac;\,a'c'}+M_{Ja,c;\,a'c'})\chi_{ac}\chi_{a'c'}^*,\quad\text{if }4|k.
  \end{cases}
\end{align*}
The action of this orbifolding, which we denote $\mathcal{O}^3$, on the minimal partition functions with $4|\overline{k}$ is given by table~\ref{tab2}.
\begin{table}[ht]
  \caption{Action of $\mathcal{O}^3$ on minimal partition functions with $4|\overline{k}$}
  \smallskip
  \centering
  \label{tab2}
  \begin{tabular}{lc}
    \toprule
    $\widetilde{M}^{2,0}$ & $(v,z,n)\leftrightarrow(v,z,n+1)$\\ \midrule
    $\widetilde{M}^{2,1}$ & $(v,z,n)\leftrightarrow(v,z,n+1)$\\ \midrule
    $\widetilde{M}^{2,2}$ & $(v,z,n,m)\leftrightarrow(v,z,n+1,m)$\\ \midrule
      $\widetilde{E}^{10}_1$ & $(6,z)\leftrightarrow(6,z)$\\ \midrule
    $\widetilde{E}^{10}_2$ & $(12,z,0,m)\leftrightarrow(12,z,0,m)$\\ \bottomrule
  \end{tabular}
\end{table}
For $\widetilde{M}^{2,0}$ and $\widetilde{M}^{2,1}$ the action coincides with that of $\mathcal{O}^1$ (as we would expect since if $\widetilde{M}_{ac;\,a'c'}\neq 0$ then $c$ is even for these families).  For $\widetilde{M}^{2,2}$ we have obtained an additional $\mZ_2$ symmetry, which along with $\mathcal{O}^1$ and $\mathcal{O}^2$ from the previous section allows us to construct an orbifolding between any two $\widetilde{M}^{2,2}$ modular invariants with $v_1=v_2$ and $z_1=\pm z_2$.  As one might expect, for the special case $v=\overline{k}$ and $z=1$ this orbifolding manifests itself as $\mathcal{A}_k\otimes\mathcal{I}_{2\overline{k}}\leftrightarrow\mathcal{D}_k\otimes\mathcal{I}_{2\overline{k}}$.  The exceptional modular invariants $\widetilde{E}^{10}_1,\widetilde{E}^{10}_2$ are left invariant.

The effect of $\mathcal{O}^3$ on the minimal models with $4|k$ is given in table~\ref{tab3}.
\begin{table}[ht]
  \caption{Action of $\mathcal{O}^3$ on minimal partition functions with $4|k$}
  \smallskip
  \centering
  \label{tab3}
  \begin{tabular}{lcl}
    \toprule
    $\widetilde{M}^{4,0}(v,z,n,m)$ & $\rightarrow$ & $\widetilde{M}^{4,2}(v,z,2m+2n+1)$\\ \midrule
    $\widetilde{M}^{4,1}(v,z,x,y)$ & $\rightarrow$ & $\widetilde{M}^{4,1}(v,z,x,y)$\\ \midrule
    $\widetilde{M}^{4,2}(v,z,x)$ & $\rightarrow$ & $\widetilde{M}^{4,2}(v,z,x)$\\ \midrule
    $\widetilde{M}^{4,3}(v,z,n)$ & $\rightarrow$ & $\widetilde{M}^{4,2}(v,z,2n+z)$\\ \midrule
    $\widetilde{E}^{16}_1(v,z,x,y)$ & $\rightarrow$ & $\widetilde{E}^{16}_1(v,z,x,y)$\\ \midrule
    $\widetilde{E}^{16}_2(v,z,x)$ & $\rightarrow$ & $\widetilde{E}^{16}_2(v,z,x)$\\ \midrule
    $\widetilde{E}^{28}(15,z,x)$ & $\rightarrow$ & $\widetilde{E}^{28}(15,z,x)$\\ \bottomrule
  \end{tabular}
\end{table}
In particular, $\widetilde{M}^{4,3}(\frac{\overline{k}}{2},1,0)=\mathcal{A}_k$ is mapped to $\widetilde{M}^{4,2}(\frac{\overline{k}}{2},1,1)=\mathcal{D}_k$ as we might expect.  The modular invariants in the families $\widetilde{M}^{4,1}$ and $\widetilde{M}^{4,2}$ and the exceptionals are left invariant.\footnote{Actually the formula given above for the $\mZ_2$ orbifolding has to be divided through by 2 in order to get $\widetilde{M}_{00;\,00}=1$.  This factor of 2 appears because $\mZ_2$ acts trivially on all the states so $Z=Z^{\text{inv}}=Z^{\text{twist}}$ and so $Z^{\text{orb}}=2Z$.}  We note that modular invariants in $\widetilde{M}^{4,0}$ and $\widetilde{M}^{4,3}$ are sent to $\widetilde{M}^{4,2}$ under this orbifolding.  This demonstrates that orbifoldings can map between, as well as within, families of minimal model partition functions.  In the next section we will show that in fact all the non-exceptional families at a given level $k$ can be mapped into one another via orbifoldings, and that the same holds true for the exceptional families.

\subsubsection{Orbifoldings between Minimal Families}\label{between}
We prove the following proposition:
\begin{proposition}\label{orbbetweenfamilies}
  \begin{enumerate}
  \item Let $4|k$.  Then all simple current invariants at level $k$ can be mapped by an orbifolding to the family $\widetilde{M}^{4,2}$.
  \item Let $4|\overline{k}$.  Then all simple current invariants at level $k$ can be mapped by an orbifolding to the family $\widetilde{M}^{2,0}$.
  \item Let $k=10$.  Then all exceptional invariants at level $k$ can be mapped by an orbifolding to the family $\widetilde{E}^{10}_1$.
  \item Let $k=16$.  Then all exceptional invariants at level $k$ can be mapped by an orbifolding to the family $\widetilde{E}^{16}_2$.
  \end{enumerate}
  (When $k$ is odd or $k=28$ there is only one family.)
\end{proposition}
We construct the necessary orbifoldings to prove statements $1$-$4$ in the following two sections.

\subsubsection{Orbifoldings between Minimal Families: \texorpdfstring{$4|k$}{4|k}}\label{O^4}
In section~\ref{O^3} we saw that the generalised $\mathcal{A}\leftrightarrow\mathcal{D}$ orbifolding $\mathcal{O}^3$ mapped members of the family $\widetilde{M}^{4,0}$ and $\widetilde{M}^{4,3}$ into the family $\widetilde{M}^{4,2}$.  We will now show that $\widetilde{M}^{4,2}$ contains an orbifold of every member of the family $\widetilde{M}^{4,1}$, and that $\widetilde{E}^{16}_2$ contains an orbifold of every member of $\widetilde{E}^{16}_1$.  This will prove parts $1$ and $4$.

Fix some $k\in 4\mZ$.  We want to construct an orbifolding which in particular sends $\widetilde{M}^{4,1}$ to $\widetilde{M}^{4,2}$.  The latter only has left-right couplings in the NS$\otimes$NS and R$\otimes$R sectors, but the former has couplings in all 4 possible sectors NS$\otimes$NS, NS$\otimes$R, R$\otimes$NS and R$\otimes$R.  In order to preserve the NS$\otimes$NS and R$\otimes$R sectors and remove the NS$\otimes$R and R$\otimes$NS sectors we define a $\mZ_2$ action by $g\cdot\chi_{ac}\chi_{a'c'}^*=(-1)^{a+c+a'+c'}\;\chi_{ac}\chi_{a'c'}^*$.  For $m,n\in\{0,1\}$ we find
\begin{align*}
  \orbox{g^m}{g^n}&=\sum_{\substack{(ac)\in Q_k\\(a'c')\in Q_k}}(-1)^{(a+c+a'+c')m}M_{J^na,c+n\overline{k};\,J^na',c'+n\overline{k}}\chi_{ac}\chi_{a'c'}^*.
\end{align*}
This transforms correctly under the $S$- and $T$-transformations, resulting in an orbifold
\begin{align*}
  Z^{\text{orb}}&=\sum_{a+c+a'+c'\equiv 0\!\!\!\!\!\mod 2}(M_{ac;\,a'c'}+M_{Ja,c+\overline{k};\,Ja',c'+\overline{k}})\chi_{ac}\chi_{a'c'}^*.
\end{align*}
We call this orbifolding $\mathcal{O}^4$.

This orbifolding acts trivially on those modular invariants which only have NS$\otimes$NS and R$\otimes$R sectors: $\widetilde{M}^{4,2}$, $\widetilde{M}^{4,3}$,  $\widetilde{E}^{16}_2$ and $\widetilde{E}^{28}$.  The action of $\mathcal{O}^4$ on the other modular invariants that occur when $4|k$ is given in table~\ref{tab4}.\footnote{Note that in the RHS of the second and fifth lines the parameter $2z$ is to be understood modulo $\frac{v^2}{2\overline{k}}$.  Recall that the $z$ parameter in each of the minimal partition functions given in section~\ref{minimalmodels} is defined modulo some integer.}
\begin{table}[ht]
  \caption{Action of $\mathcal{O}^4$ on minimal partition functions with $4|k$}
  \smallskip
  \centering
  \label{tab4}
  \begin{tabular}{lcl}
    \toprule
    $\widetilde{M}^{4,0}(v,z,n,m)$ & $\rightarrow$ & $\widetilde{M}^{4,2}(v,z,2m+2n+1)$\\ \midrule
    $\widetilde{M}^{4,1}(v,z,x,y)$ & $\rightarrow$ & $\widetilde{M}^{4,2}(\frac{v}{2},2z,y-x+1)$\\ \midrule
    $\widetilde{M}^{4,2}(v,z,x)$ & $\rightarrow$ & $\widetilde{M}^{4,2}(v,z,x)$\\ \midrule
    $\widetilde{M}^{4,3}(v,z,n)$ & $\rightarrow$ & $\widetilde{M}^{4,3}(v,z,2n+z)$\\ \midrule
    $\widetilde{E}^{16}_1(v,z,x,y)$ & $\rightarrow$ & $\widetilde{E}^{16}_2(\frac{v}{2},2z,y-x+1)$\\ \midrule
    $\widetilde{E}^{16}_2(v,z,x)$ & $\rightarrow$ & $\widetilde{E}^{16}_2(v,z,x)$\\ \midrule
    $\widetilde{E}^{28}(15,z,x)$ & $\rightarrow$ & $\widetilde{E}^{28}(15,z,x)$\\ \bottomrule
  \end{tabular}
\end{table}

\subsubsection{Orbifoldings between Minimal Families: \texorpdfstring{$4|\overline{k}$}{4|kbar}}\label{O^5andO^6}
In this section we shall show that all non-exceptional invariants with $4|\overline{k}$ can be sent into $\widetilde{M}^{2,0}$ by an orbifolding, and all exceptional invariants with $k=10$ can be sent into $\widetilde{E}^{10}_1$, proving parts $2$ and $3$ of proposition~\ref{orbbetweenfamilies}.

First we shall construct an orbifolding $\mathcal{O}^5$ from $\widetilde{M}^{2,1}$ to $\widetilde{M}^{2,0}$.  Fix a $k$ with $4|\overline{k}$ and fix $(v,z,n)$ satisfying $\frac{\overline{k}}{2v}\in\mZ$, $\frac{2v^2}{\overline{k}}\in 2\mZ+1$ and $\frac{\overline{k}(z^2-1)}{2v^2}\in\mZ$ where $z\in\{1,\ldots,\frac{2v^2}{\overline{k}}\}$ and $n\in\{0,1\}$.  Then from section~\ref{minimalmodels} there is a minimal partition function $\widetilde{M}^{2,1}(v,z,n)$.  We need to define a group action on the states of $M\equiv\widetilde{M}^{2,1}(v,z,n)$.  Note that $M_{ad;\,a'd'}\neq 0\Rightarrow d=\frac{c\overline{k}}{2v},\; d'=\frac{c'\overline{k}}{2v}$ and $c+c'\equiv 0\!\!\mod 2$; thus there is a $\mZ_2$ action on the states given by $g\cdot\chi_{a,\frac{c\overline{k}}{2v}}\chi_{a,\frac{c'\overline{k}}{2v}}^*=(-1)^{\frac{c+c'}{2}}\chi_{a,\frac{c\overline{k}}{2v}}\chi_{a,\frac{c'\overline{k}}{2v}}^*$ and which for $n,m\in\{0,1\}$ gives rise to
\begin{align*}
  \orbox{g^m}{g^n}&=\sum_{\substack{(ac)\in Q_k\\(a'c')\in Q_k}}M_{ac;\,a'c'}\;e^{\frac{2\pi imv}{\overline{k}}\left(\frac{c+c'}{2}\right)}\chi_{a,c-nv}\chi_{a,c'+nv}^*,
\end{align*}
whence we conclude that
\begin{align*}
  Z^{\text{orb}}&=\sum_{\substack{(ac)\in Q_k\\(a'c')\in Q_k\\}}\left(M_{ac;\,a'c'}+M_{a,c+v;\,a',c-v}\right)\;\delta(c+c'\equiv0\mod\frac{2\overline{k}}{v})\;\chi_{ac}\chi_{a'c'}^*.
\end{align*}
Inserting $M=M^{2,1}(v,z,n)$ from equation~\eqref{M^{2,0}} one finds $Z^{\text{orb}}=\widetilde{M}^{2,0}(v',z',n)$, where $v'=2v$, $z'=\left(\frac{2v^2}{\overline{k}}\right)^2(3-z)$, where we understand $z'$ to be defined modulo $\frac{8v^2}{\overline{k}}$.  We have therefore demonstrated that every model with partition function in $\widetilde{M}^{2,1}$ gives rise to a $\mZ_2$ orbifold in $\widetilde{M}^{2,0}$.

Constructing an orbifolding $\mathcal{O}^6$ from $\widetilde{M}^{2,2}$ to $\widetilde{M}^{2,0}$ is similar:  fixing some $k$ such that $4|\overline{k}$, we define a $\mZ_2$ action by $g\cdot\chi_{ac}\chi_{a'c'}^*=(-1)^c\chi_{ac}\chi_{a'c'}^*$.  We claim that for $m,n\in\{0,1\}$
\begin{align*}
  \orbox{g^m}{g^n}&=\sum_{\substack{(ac)\in Q_k\\(a'c')\in Q_k}}(-1)^{cm}M_{a,c+n\overline{k};\,a'c'}\chi_{ac}\chi_{a'c'}^*.
\end{align*}
This is evidently correct when $n=0$ and it is not hard to check that it transforms correctly under the $S$ and $T$ transformations.  It yields
\begin{align}
  Z^{\text{orb}}&=\sum_{\substack{(ac)\in Q_k\\(a'c')\in Q_k}}\left[M_{ac;\,a'c'}+M_{a,c+\overline{k};\,a'c'}\right]\delta(c\equiv0\!\!\!\mod2)\chi_{ac}\chi_{a'c'}^*.\label{O^6equation}
\end{align}
Choosing some $v,z$ such that $\frac{\overline{k}}{v}$ is odd and $\frac{v^2}{\overline{k}},\frac{\overline{k}(z^2-1)}{4v^2}\in\mZ$, we can apply $\mathcal{O}^6$ to the modular invariant $M\equiv\widetilde{M}^{2,2}(v,z,n,m)$.  Using equation~\eqref{M^{2,2}} and~\eqref{M^{2,0}} we find
\begin{align*}
  Z^{\text{orb}}&=\widetilde{M}^{2,0}(v',z,n)
\end{align*}
where we have set $2v'=v$ and $z$ is now understood to be defined modulo $\frac{2v'^2}{\overline{k}}$.

It remains to show that the family $\widetilde{E}^{10}_2$ can be mapped via an orbifolding into the family $\widetilde{E}^{10}_1$.  We simply apply the orbifolding $\mathcal{O}^6$ from the previous section to the exceptional invariant $\widetilde{E}^{10}_2(12,v,0,m)$: substituting~\eqref{E^{10}_2} into~\eqref{O^6equation} we obtain
\begin{align*}
  Z^{\text{orb}}&=\widetilde{E}^{10}_1(6,z).
\end{align*}
This completes the proof of proposition~\ref{orbbetweenfamilies}.\qed

\subsubsection{Orbifoldings within Minimal Families -- a Useful Formula}\label{usefulformula}
In order to complete the proof of theorem~\ref{theorem1}, we must find orbifoldings within the families $\widetilde{M}^0$, $\widetilde{M}^{4,2}$, $\widetilde{M}^{2,0}$, $\widetilde{E}^{10}_1$, $\widetilde{E}^{16}_2$ and $\widetilde{E}^{28}$ which map all members down to a specific partition function.  Since we already have control of the $\mZ_2$ parameters (labelled by $n$ or $x$) via the orbifoldings $\mathcal{O}^1$ and $\mathcal{O}^2$, in this section we concentrate on trying to control the parameters $v$ and $z$.

We begin by considering a general orbifolding by a group $\mZ_{\beta}$, acting on the $\mathfrak{u}(1)$ label $c$ on the left-hand side.\footnote{Note that the remaining families are all symmetric, so it doesn't matter whether we act on the left- or right-hand sides.}  Fix a modular invariant $M$ in one of the above families and take the largest integer $\alpha$ such that
\begin{align*}
  M_{ac;\,a'c'}\neq 0&\Rightarrow c,c'\in\alpha\mZ.
\end{align*}
For these families, $\frac{\overline{k}}{\alpha^2}\in\mZ$.  We will define a $\mZ_{\beta}$-orbifolding $\mathcal{O}^7$ for some integer $\beta$ satisfying $\beta|\frac{\overline{k}}{\alpha^2}$.  Let $\mZ_{\beta}=\langle g\rangle$ act on the states of $M$ via
\begin{align*}
  g\cdot\chi_{a,\alpha c}\chi_{a',\alpha c'}^*=e^{\frac{2\pi ic}{\beta}}\chi_{a,\alpha c}\chi_{a',\alpha c'}^*.
\end{align*}
We claim that the result is
\begin{align*}
  \orbox{g^m}{g^n}&=\sum_{\substack{a,a'=0,\ldots,k\\c,c'\in\mZ_{\frac{2\overline{k}}{\alpha}}}}M_{a,\alpha c;\,a',\alpha c'}\,\,e^{\frac{2\pi im}{\beta}\left(c-\frac{n\overline{k}}{\alpha^2\beta}\right)}\chi_{a,\alpha\left(c-\frac{2n\overline{k}}{\alpha^2\beta}\right)}\chi_{a',\alpha c'}^*.
\end{align*}
It is easy to see this is correct when $n=0$.  One then checks that it behaves correctly under the action of the $S$- and $T$-transformations.  The line of attack is the usual one:  for the $T$-transformation we use the integer-spin condition (equation~\eqref{Tcondition}) to remove the otherwise unwieldy factor of $e^{2\pi i(h_{ac}-h_{a'c'})}$; and for the $S$-matrix we use the nice behaviour of the simple current action (equation~\eqref{simplecurrentaction}) to juggle unwanted factors on and off the $S$-matrices until one has something of the form $SMS^{\dagger}$, which can be replaced with $M$, just as we did in section~\ref{O^1}.\footnote{Fans of dense technical details will be pleased to learn that all the calculations alluded to in this section are written out explicitly in~\cite{Gray2009}.}

The partition function of the orbifolding $\mathcal{O}^7$ is then given by the sum over the twisted sectors:
\begin{align}
  Z^{\text{orb}}&=\sum_{\substack{N=0,\ldots,\beta-1\\a=0,\ldots,k\\a'=0,\ldots,k}}\,\sum_{\substack{s\in\mZ_{\frac{2\overline{k}}{\alpha\beta}}\\c'\in\mZ_{\frac{2\overline{k}}{\alpha}}}}M_{a,\alpha\left(s\beta+\frac{N\overline{k}}{\alpha^2\beta}\right);\,a',\alpha c'}\chi_{a,\alpha\left(s\beta-\frac{N\overline{k}}{\alpha^2\beta}\right)}\chi_{a',\alpha c'}^*.\label{O^7}
\end{align}
If it happens that $\frac{\overline{k}}{\alpha^2\beta^2}\in\mZ$ then the above simplifies to
\begin{align}
  Z^{\text{orb}}&=\sum_{\substack{N=0,\ldots,\beta-1\\a=0,\ldots,k\\a'=0,\ldots,k}}\,\sum_{\substack{c\in\mZ_{\frac{2\overline{k}}{\alpha\beta}}\\c'\in\mZ_{\frac{2\overline{k}}{\alpha}}}}M_{a,\alpha\beta\left(c+\frac{2N\overline{k}}{\alpha^2\beta^2}\right);\,a',\alpha c'}\chi_{a,\alpha\beta c}\chi_{a',\alpha c'}^*.\label{O^7simple}
\end{align}

\subsubsection{Controlling the Parameter \texorpdfstring{$v$}{v}}\label{vcontrol}
The aim of this section is to find an orbifolding which sends the parameter $v$ to the smallest possible value it can take:
\begin{proposition}\label{smallestv}
  Fix $k$ and let $M$ be a level $k$ modular invariant in one of the families $\widetilde{M}^0$, $\widetilde{M}^{4,2}$, $\widetilde{M}^{2,0}$, $\widetilde{E}^{10}_1$, $\widetilde{E}^{16}_2$ or $\widetilde{E}^{28}$ with parameters $(v,z,*)$ where $*$ is either $n$ or $x$.  Then we can map $M$ via an orbifolding to a minimal partition function in the same family with parameters $(v',z,*)$ where $v'$ is the smallest possible value of $v$ allowed.
\end{proposition}
In the exceptional cases $\widetilde{E}^{10}_1$ and $\widetilde{E}^{28}$ there is only one allowed value of $v$, so the proposition is trivial in these cases; they are included for completeness.

We shall prove the claim using the orbifoldings constructed in section~\ref{usefulformula}.  The idea is to map by the orbifolding with the largest possible value of $\beta$ that satisfies $\frac{\overline{k}}{\alpha^2\beta^2}\in\mZ$.  Again we recommend the eager to read~\cite{Gray2009} for the full computations.

\subsubsection{\texorpdfstring{$k$}{k} odd}
Let $k$ be an odd integer and let $M$ be a modular invariant at level $k$ with parameters $(v,z,n)$ (see~\eqref{M^0}).  Write $\overline{k}=\prod_{i=1}^lp_i^{2a_i+\delta_i}$ where the $p_i$ are distinct odd primes and $\delta_i\in\{0,1\}$ for each $i=1,\ldots,l$.  Similarly write $v=\prod_{i=1}^lp_i^{b_i}$ for some integers $b_i$.  The conditions $\frac{\overline{k}}{v},\frac{v^2}{\overline{k}}\in\mZ$ are equivalent to $a_i+\delta_i\leq b_i\leq 2a_i+\delta_i$, so we can define an integer $\beta=\prod_{i=1}^lp_i^{b_i-a_i-\delta_i}$.

As in the previous section we find the biggest integer $\alpha$ such that $M_{ac;\,a'c'}\neq0\Rightarrow c,c'\in\alpha\mZ$; here, $\alpha=\frac{\overline{k}}{v}=\prod_{i=1}^lp_i^{2a_i-b_i+\delta_i}$.  With these values we see that $\frac{\overline{k}}{\alpha^2\beta^2}=\prod_{i=1}^lp_i^{\delta_i}\in\mZ$, so we can perform $\mathcal{O}^7$, the $\mZ_{\beta}$ orbifolding from the previous section, on $M$ using the simplified formula in equation~\eqref{O^7simple}.  After a page of computation we arrive at
\begin{align*}
  Z^{\text{orb}}&=\widetilde{M}^0(v',z,n),
\end{align*}
where we have defined $v'=\frac{\overline{k}}{\alpha\beta}=\prod_{i=1}^lp_i^{a_i+\delta_i}$.  Note that this is the smallest divisor $v'$ of $\overline{k}$ satisfying $\frac{v'^2}{\overline{k}}\in\mZ$.  Thus we have successfully minimised the parameter $v$.

\subsubsection{\texorpdfstring{$4$}{4} divides \texorpdfstring{$k$}{k}}
The $\widetilde{M}^{4,2}$ case is similar.  Fix $k$ such that $4|k$ and choose an $\widetilde{M}^{4,2}$ modular invariant with parameters $(v,z,x)$.  We write $\overline{k}=2\prod_{i=1}^lp_i^{2a_i+\delta_i}$ with $p_i$ distinct odd primes and $\delta_i\in\{0,1\}$ and write $v=\prod_{i=1}^lp_i^{b_i}$ for some integers $b_i$.  This time $\alpha=\frac{\overline{k}}{2v}=\prod_{i=1}^lp_i^{2a_i-b_i+\delta_i}$ and we set $\beta=\prod_{i=1}^lp_i^{b_i-a_i-\delta_i}$.  Again we find that $\frac{\overline{k}}{\alpha^2\beta^2}=\prod_{i=1}^lp_i^{\delta_i}\in\mZ$ so we can apply equation~\eqref{O^7simple} to the partition function given by equations~\eqref{M^{4,2}} in order to calculate the $\mZ_{\beta}$ orbifolding.  The result is
\begin{align*}
  Z^{\text{orb}}&=\widetilde{M}^{4,2}(v',z,x),
\end{align*}
where we have defined $v'=\frac{\overline{k}}{2\alpha\beta}=\prod_{i=1}^lp_i^{a_i+\delta_i}$.  This shows that for a fixed $k$ we can always send $v$ to its smallest possible value in the family $\widetilde{M}^{4,2}$.

\subsubsection{\texorpdfstring{$4$}{4} divides \texorpdfstring{$\overline{k}$}{kbar}}\label{vcontrolend}
Finally we address the case when $k$ satisfies $4|\overline{k}$.  Fix such a $k$ and a $\widetilde{M}^{2,0}$ modular invariant $M$ with parameters $(v,z,n)$ (see equation~\eqref{M^{2,0}}).  As before write $\overline{k}=\prod_{i=0}^lp_i^{2a_i+\delta_i}$ where $p_0=2$ and the $p_i$ are distinct odd primes for $i\geq 1$, $\delta_i\in\{0,1\}$ for each $i=0,\ldots,l$ and $a_0\geq 1$.  For this partition function $\alpha=\frac{\overline{k}}{v}=\prod_{i=0}^lp_i^{2a_i+\delta_i-b_i}$ and we set $\beta=\prod_{i=0}^lp_i^{b_i-a_i-\delta_i}$, which is bound to be an integer by the condition $\frac{v^2}{\overline{k}}\in\mZ$.  We find once again that $\frac{\overline{k}}{\alpha^2\beta^2}=\prod_{i=0}^lp_i^{\delta_i}\in\mZ$ and so we can use the formula~\eqref{O^7simple} to calculate the $\mZ_{\beta}$ orbifold of $M$.  Substituting in equation~\eqref{M^{2,0}} we find
\begin{align*}
  Z^{\text{orb}}&=\widetilde{M}^{2,0}(v',z,n)
\end{align*}
where we have defined $v'=\frac{\overline{k}}{\alpha\beta}=\prod_{i=0}^lp_i^{a_i+\delta_i}$.  This completes the proof of proposition~\ref{smallestv} for the simple current invariants.

It remains to check the case $\widetilde{E}^{16}_2$.  Let $M$ be the modular invariant in $\widetilde{E}^{16}_2$ with parameters $(v=9,z,x)$.  Then $\alpha=1$ and we choose $\beta=3$ so that $\frac{\overline{k}}{\alpha^2\beta^2}=2\in\mZ$.  It is then straight-forward to apply equation~\eqref{O^7simple} to find
\begin{align*}
  Z^{\text{orb}}&=\widetilde{E}^{16}_2(3,1,x).
\end{align*}
This completes the proof of proposition~\ref{smallestv}.\qed

\subsubsection{Controlling the Parameter \texorpdfstring{$z$}{z}}\label{zcontrol}
Now that we can map via orbifoldings any minimal partition function into a particular family with a particular value of $v$, it remains to find an orbifolding which lets us control the parameter $z$.  We will prove
\begin{proposition}\label{z=1}
  Fix $k$ and let $M$ be a level $k$ modular invariant in one of the families $\widetilde{M}^0$, $\widetilde{M}^{4,2}$, $\widetilde{M}^{2,0}$, $\widetilde{E}^{10}_1$, $\widetilde{E}^{16}_2$ or $\widetilde{E}^{28}$ with parameters $(v,z,*)$ where $v$ is as small as possible and $*$ is either $n$ or $x$.  Then we can map $M$ via orbifoldings to a minimal partition function in the same family with parameters $(v,z',*)$ where
  \begin{align*}
    2z\equiv1\mod\frac{v^2}{\overline{k}} & \text{ for odd $k$},\\
    z\equiv1\mod\frac{2v^2}{\overline{k}} & \text{ otherwise.}
  \end{align*}
\end{proposition}
When $v$ is minimised in the family $\widetilde{E}^{16}_2$ then $z$ is forced to be $1$, so the statement is trivial in this case; it is included in the proposition only for completeness.

\subsubsection{\texorpdfstring{$k$}{k} odd}
Let $k$ be odd and let $M$ be a level $k$ modular invariant with parameters $(v,z,n)$ where $v$ is as small as possible (see equation~\eqref{M^0}).  Write $\overline{k}=\prod_{i=1}^lp_i^{2a_i+1}\times\prod_{j=1}^mq_j^{2b_j}$ where the $p_i$ and $q_j$ are mutually distinct odd primes.  Then we must have $v=\prod_{i=1}^lp_i^{a_i+1}\prod_{j=1}^mq_j^{b_j}$, since $v$ is the smallest solution to $\frac{\overline{k}}{v},\frac{v^2}{\overline{k}}\in\mZ$, and therefore $\frac{v^2}{\overline{k}}=\prod_{i=1}^lp_i$.  Now $z$ is defined to be a solution to $4z^2-1\equiv0\mod\frac{v^2}{\overline{k}}$.  So we have
\begin{align*}
  (2z+1)(2z-1)&\equiv0\mod\prod_{i=1}^lp_i.
\end{align*}
But since a given odd prime cannot divide both $2z+1$ and $2z-1$, is it equivalent to say that there must exist a partition $\{p_{i_1},\ldots,p_{i_t}\}\cup\{p_{j_1},\ldots,p_{j_u}\}$ of the $p_i$ such that
\begin{align*}
  \begin{cases}
    2z+1\equiv0\mod\prod_{k=1}^{t}p_{i_k},\\
    2z-1\equiv0\mod\prod_{k=1}^{u}p_{j_k}.
  \end{cases}
\end{align*}
We are trying to map this partition function via an orbifolding to one where $z$ is given by the choice of partition $\{\}\cup\{p_1,\ldots,p_l\}$.  So we set $\beta=\prod_{k=1}^{t}p_{i_k}$ and try to make a $\mZ_{\beta}$ orbifold.  Recall that the largest integer $\alpha$ satisfying the condition
\begin{align*}
  M_{ac;\,a'c'}\neq0&\Rightarrow c,c'\in\alpha\mZ
\end{align*}
is $\alpha=\frac{\overline{k}}{v}=\prod_{i=1}^lp_i^{a_i}\prod_{j=1}^mq_j^{b_j}$.  Thus $\frac{\overline{k}}{\alpha^2\beta}=\prod_{k=1}^up_{j_k}\in\mZ$ and we can apply the orbifolding in equation~\eqref{O^7}.\footnote{Again, computations can be found in~\cite{Gray2009}.}
\begin{align*}
  Z^{\text{orb}}&=\widetilde{M}^0(v,z',n)
\end{align*}
where $z'$ is the unique solution to $2z\equiv 1$ modulo $\frac{v^2}{\overline{k}}$ as required.

\subsubsection{\texorpdfstring{$4$}{4} divides \texorpdfstring{$k$}{k}}
The proof of proposition~\ref{z=1} in the case where $4|k$ proceeds in a very similar way to the case where $k$ is odd.  Fix a modular invariant $M\equiv\widetilde{M}^{4,2}$ with parameters $(v,z,x)$ where $v$ is minimal.  Write $\overline{k}=2\prod_{i=1}^lp_i^{2a_i+1}\prod_{j=1}^mq_j^{2b_j}$ with $p_i,q_j$ mutually distinct odd primes and note that since $v$ is minimal (see equation~\eqref{M^{4,2}}) we must have $v=\prod_{i=1}^lp_i^{a_i+1}\prod_{j=1}^mq_j^{b_j}$ and $\frac{2v^2}{\overline{k}}=\prod_{i=1}^lp_i$.  The equation for $z$ for $\widetilde{M}^{4,2}$ is $z^2-1\equiv0\!\!\mod\frac{2v^2}{\overline{k}}$ so we have $(z+1)(z-1)\equiv0\mod\prod_{i=1}^lp_i$.  Equivalently, there exists a $t$ such that, after relabelling the $p_i$,
\begin{align*}
  \begin{cases}
    z+1\equiv0\mod\prod_{i=1}^tp_i,\\
    z-1\equiv0\mod\prod_{i=t+1}^lp_i.
  \end{cases}
\end{align*}
This time $\alpha=\frac{\overline{k}}{2v}=\prod_{i=1}^lp_i^{a_i}\prod_{j=1}^mq_j^{b_j}$ and again we set $\beta=\prod_{i=1}^tp_i$.  Then we can perform the $\mZ_{\beta}$ orbifolding given in equation~\eqref{O^7} on $M$.  This end result is
\begin{align*}
  Z^{\text{orb}}&=\widetilde{M}^{4,2}(v,1,x)
\end{align*}
as required.

\subsubsection{\texorpdfstring{$4$}{4} divides \texorpdfstring{$\overline{k}$}{kbar}}
The case where $4$ divides $\overline{k}$ is again very similar.  Fix a modular invariant $M\equiv\widetilde{M}(v,z,n)$ where $v$ is minimal.  We write $\overline{k}$ in the form $\overline{k}=2^{2r+\epsilon}\prod_{i=1}^lp_i^{2a_i+1}\times\prod_{j=1}^mq_j^{2b_j}$ with $p_i,q_j$ mutually distinct odd primes, $r\geq 1$ and $\epsilon\in\{0,1\}$.  Note that since $v$ is minimal (see~\eqref{M^{2,0}}) we must have $v=2^{r+\epsilon}\prod_{i=1}^lp_i^{a_i+1}\prod_{j=1}^mq_j^{b_j}$ and $\frac{2v^2}{\overline{k}}=2^{1+\epsilon}\prod_{i=1}^lp_i$.  Since $z$ satisfies $z^2-1\equiv0\!\!\mod\frac{2v^2}{\overline{k}}$ we must have $(z+1)(z-1)\equiv0\mod2^{1+\epsilon}\prod_{i=1}^lp_i$.  Equivalently, there exists a $t$ such that, after relabelling the $p_i$,
\begin{align*}
  \begin{cases}
    z+1\equiv0\mod2\prod_{i=1}^tp_i,\\
    z-1\equiv0\mod2\prod_{i=t+1}^lp_i.
  \end{cases}
\end{align*}
We have $\alpha=\frac{\overline{k}}{v}=2^r\prod_{i=1}^lp_i^{a_i}\prod_{j=1}^mq_j^{b_j}$ and we set $\beta=2^{x\epsilon}\prod_{i=1}^tp_i$ where $x$ is either 0 or 1 and will be specified later.  Then $\frac{\overline{k}}{\alpha^2\beta}=2^{\epsilon(1-x)}\prod_{i=t+1}^lp_i$ is an integer, so we may perform the $\mZ_{\beta}$ orbifolding given in equation~\eqref{O^7} on $M$:
\begin{align*}
  Z^{\text{orb}}&=\widetilde{M}^{2,0}(v,1,n)
\end{align*}
which completes the proof of proposition~\ref{z=1} for the simple current invariants.

\subsubsection{The Exceptional Cases}\label{zcontrolend}
When $k=10$ we need to show that there is an orbifolding connecting the $\widetilde{E}^{10}_1$ invariants with those with parameters $(v=6,z=5)$ and $(v=6,z=1)$.  But we have already seen in table~\ref{O^2table} that the orbifolding $\mathcal{O}^2$ acts on $\widetilde{E}^{10}_1(6,z)$ by $z\leftrightarrow-z\mod6$.

When $k=28$ we follow exactly the method we used for the simple current invariants when $4|k$:  we have $\overline{k}=30=2\cdot3\cdot5$ and $v=15$.  The solutions to $z^2-1\equiv0\mod15$ are $z\in\{1,4,11,14\}$ (see equation~\eqref{E^{28}}), corresponding respectively to the situations 
\begin{align*}
  z=1, & \begin{Bmatrix}
    z+1\equiv0\!\!\mod1\\
    z-1\equiv0\!\!\mod15
  \end{Bmatrix}, & \beta=1\\
  z=4, & \begin{Bmatrix}
    z+1\equiv0\!\!\mod5\\
    z-1\equiv0\!\!\mod3
  \end{Bmatrix}, & \beta=5\\
  z=11, & \begin{Bmatrix}
    z+1\equiv0\!\!\mod3\\
    z-1\equiv0\!\!\mod5
  \end{Bmatrix}, & \beta=3\\
  z=14, & \begin{Bmatrix}
    z+1\equiv0\!\!\mod15\\
    z-1\equiv0\!\!\mod1
  \end{Bmatrix}, & \beta=15.\\
\end{align*}
In each case $\alpha=1$ and so we apply orbifolding $\mathcal{O}^7$ to the invariants $M\equiv\widetilde{E}^{28}(15,z,x)$ using equation~\eqref{O^7}.  The end result is
\begin{align*}
  Z^{\text{orb}}&=\widetilde{E}^{28}(15,1,x).
\end{align*}
This completes the proof of proposition~\ref{z=1}.\qed

\subsubsection{Proof of the Theorem}\label{summary}
We are now ready to prove theorem~\ref{theorem1}.  We will restate the theorem here in a little more detail.  For notation, see section~\ref{minimalmodels}.
\begin{theorem}[Reformulation of theorem~\ref{theorem1}]\label{theorem2}
  \mbox{}
  \begin{itemize}
  \item  Let $k$ be odd and let $M$ be a simple current invariant at level $k$.  Then there exists a chain of orbifoldings mapping $M$ to $\mathcal{A}_k\otimes\overline{M}$ where $\mathcal{A}_k$ is the diagonal $\mathfrak{su}(2)$ invariant at level $k$ and the non-zero values of $\overline{M}$ are given by
    \begin{align*}
      \overline{M}_{\frac{c\overline{k}}{v},\frac{c'\overline{k}}{v}}=1\iff c'\equiv c\!\!\!\mod \frac{2v^2}{\overline{k}}
    \end{align*}
    where $v$ is the smallest divisor of $\overline{k}$ satisfying $\frac{v^2}{\overline{k}}\in\mZ$.
  \item  Let $4|\overline{k}$ and let $M$ be a simple current invariant at level $k$.  Then there exists a chain of orbifoldings mapping $M$ to $\mathcal{A}_k\otimes\overline{M}$ where $\mathcal{A}_k$ is the diagonal $\mathfrak{su}(2)$ invariant at level $k$ and the non-zero values of $\overline{M}$ are given by
    \begin{align*}
      \overline{M}_{\frac{c\overline{k}}{v},\frac{c'\overline{k}}{v}}=1\iff c'\equiv c\!\!\!\mod \frac{2v^2}{\overline{k}}
    \end{align*}
    where $v$ is the smallest divisor of $\frac{\overline{k}}{2}$ satisfying $\frac{v^2}{\overline{k}}\in\mZ$.
  \item  Let $4|k$ and let $M$ be a simple current invariant at level $k$.  Then there exists a chain of orbifoldings mapping $M$ to $\mathcal{D}_k\otimes\overline{M}$ where $\mathcal{D}_k$ is the level $k$ $\mathcal{D}$ invariant in the $\mathfrak{su}(2)$ \cADE classification, and the non-zero values of $\overline{M}$ are given by
    \begin{align*}
      \overline{M}_{\frac{c\overline{k}}{2v},\frac{c'\overline{k}}{2v}}=1\iff c'\equiv c\!\!\!\mod \frac{8v^2}{\overline{k}}
    \end{align*}
    where $v$ is the smallest divisor of $\frac{\overline{k}}{2}$ satisfying $\frac{2v^2}{\overline{k}}\in\mZ$.
  \item  Let $M$ be an exceptional invariant at level $k=10$.  Then there exists a chain of orbifoldings mapping $M$ to $\mathcal{E}^{10}\otimes\overline{M}$ where $\mathcal{E}^{10}$ is the exceptional $\mathfrak{su}(2)$ invariant at level $10$ and the non-zero values of $\overline{M}$ are given by
    \begin{align*}
      \overline{M}_{2c,2c'}=1\iff c'\equiv c\!\!\!\mod 6.
    \end{align*}
  \item  Let $M$ be an exceptional invariant at level $k=16$.  Then there exists a chain of orbifoldings mapping $M$ to $\mathcal{E}^{16}\otimes\overline{M}$ where $\mathcal{E}^{16}$ is the exceptional $\mathfrak{su}(2)$ invariant at level $16$ and the non-zero values of $\overline{M}$ are given by
    \begin{align*}
      \overline{M}_{3c,3c'}=1\iff c'\equiv c\!\!\!\mod 4.
    \end{align*}
  \item  Let $M$ be an exceptional invariant at level $k=28$.  Then there exists a chain of orbifoldings mapping $M$ to $\mathcal{E}^{28}\otimes\overline{M}$ where $\mathcal{E}^{28}$ is the exceptional $\mathfrak{su}(2)$ invariant at level $28$ and $\overline{M}$ is given by
    \begin{align*}
      \overline{M}_{c,c'}=1\iff c'\equiv c\!\!\!\mod 60.
    \end{align*}
  \end{itemize}
\end{theorem}
\begin{proof}
  The requisite orbifoldings were constructed in the preceding sections.  Given a modular invariant $M$ at level $k$, we use proposition~\ref{orbbetweenfamilies} (if necessary) to map $M$ into one of the families $\widetilde{M}^0,\widetilde{M}^{2,0},\widetilde{M}^{4,2},\widetilde{E}^{10}_1,\widetilde{E}^{16}_2$ or $\widetilde{E}^{28}$, uniquely determined by the value of $k$ and whether $M$ is a simple current invariant or an exceptional invariant.  We can then apply proposition~\ref{smallestv} to map the parameter  $v$ to the smallest possible value it can take for the given $k$, while leaving the other parameters unchanged.  Proposition~\ref{z=1} sends $z$ to 1 if $k$ is even, and sets $2z\equiv1$ if $k$ is odd.  Finally, if necessary, we use the orbifolding $\mathcal{O}^1$ of section~\ref{O^1} to fix $n=0$ when $k$ is odd or $4|\overline{k}$; or to fix $x=1$ when $4|k$.  The resulting partition functions are given explicitly above using equations~\eqref{M^0}--\eqref{E^{28}}.\qed
\end{proof}

\section{Analysis of the Simple Current Invariants}
\subsection{The Kreuzer-Schellekens Construction}
  In~\cite{Kreuzer1994} it is shown that all simple current invariants which obey both $1$-loop and higher-genus modular invariance can be obtained as orbifolds of the diagonal modular invariant by a subgroup of the centre.  It is conjectured that all simple current modular invariants can be obtained in this way; that is, it is conjectured that the constraint of higher-genus modular invariance is in fact superfluous.  We will analyse the solutions of Gannon's classification to show that this is indeed the case for the unitary $N=2$ minimal models.

\subsubsection{\texorpdfstring{$k$}{k} odd}
One can easily read off from Gannon's classification that every modular invariant with $k$ odd is a simple current invariant.  Furthermore, following~\cite{Kreuzer1994}, precisely one modular invariant can be constructed as an orbifold for each subgroup of the effective centre $\mathcal{C}\cong\mZ_{2\overline{k}}$ (there is no discrete torsion in this case, since subgroups of $\mZ_{2\overline{k}}$ are cyclic).

One can check using induction on the number of prime factors that the number of subgroups of $\mZ_{q}$, equal to the number of divisors of $q$, is $d(q):=\prod_{i=1}^l(1+n_i)$ where $q$ is written $q=\prod_{i=1}^lp_i^{n_i}$ for distinct primes $p_i$.  The following lemma establishes that the number of modular invariants at each odd level $k$ (see equation~\eqref{M^0}) is precisely the number of subgroups of $\mZ_{2\overline{k}}$, showing that the Schellekens-Kreuzer orbifold construction does indeed give all modular invariants when the level $k$ is odd.

\begin{lemma}
  Let $k$ be odd.  Then the number of solutions $(v,z,n)\in\{1,\ldots,\overline{k}\}\times\{1,\ldots,\frac{v^2}{\overline{k}}\}\times\{0,1\}$ to the equations
  \begin{align*}
    \begin{matrix}
      \frac{v^2}{\overline{k}},\frac{\overline{k}}{v}\in\mZ, & 4z^2\equiv 1\!\!\!\mod \frac{v^2}{\overline{k}}
    \end{matrix}
  \end{align*}
  is equal to $d(2\overline{k})$.
\end{lemma}
The proof is a simple counting argument.  The main step is counting the number of possible values of $z$ for a given $v$, and we partially solved this problem already in constructing the $z$-controlling orbifoldings of section~\ref{zcontrol}.  For a detailed proof, we refer the reader to the author's PhD thesis~\cite{Gray2009}.\qed

\subsubsection{4 divides \texorpdfstring{$k$}{k}}
We now turn our attention to the case when $4|k$.  Again we can immediately read off from Gannon's classification that $\widetilde{M}^{4,0}$, $\widetilde{M}^{4,1}$, $\widetilde{M}^{4,2}$ and $\widetilde{M}^{4,3}$ are all simple current invariants.

The subgroups of the effective centre $\mathcal{C}_k\cong\mZ_2\times\mZ_{2\overline{k}}$ are given by
\begin{align*}
  \begin{matrix}
    \mZ_2\times\mZ_l\cong\mZ_{2l},&2l|\overline{k}\\
    \mZ_2\times\mZ_{2l},&l|\overline{k}\\
    \{0\}\times\mZ_{l}\cong\mZ_{l},&l|2\overline{k}\\
    \langle(J,\frac{\overline{k}}{l})\rangle\cong\mZ_{2l},&l|\overline{k}.
  \end{matrix}
\end{align*}
We can define an orbifold for each subgroup of the centre and for each choice of discrete torsion associated to that subgroup.  For a cyclic group $\mZ_q$ there is no choice to make; for a group $\mZ_2\times\mZ_{2q}$ there are two degrees of freedom.  Writing $\tau(G)$ for the number of degrees of freedom coming from discrete torsion associated to the group $G$, we find the number of simple current invariants obtained via an orbifold of the diagonal invariant when $4|k$ is
\begin{align*}
  N&=\sum_{G\leq \mZ_2\times\mZ_{2\overline{k}}}\tau(G)=5d(\overline{k})
\end{align*}
where $d(q)$, as above, is the number of divisors of $q$.

The following lemma shows that if $4|k$ then the number of simple current modular invariants is equal to $N=5d(\overline{k})$, the number of orbifolds of the diagonal invariant, so the Schellekens-Kreuzer construction does again find all simple currents invariants when $4|k$.

\begin{lemma}
  Let $8|k+4$.  Then the number of solutions $(v,z,n,m)\!\in\!\{1,\ldots,\frac{\overline{k}}{2}\}\times\{1,\ldots,\frac{2v^2}{\overline{k}}\}\times\{0,1\}^2$ to the equations
  \begin{align*}
    \begin{matrix}
      \frac{2v^2}{\overline{k}},\frac{\overline{k}}{2v}\in\mZ, & z^2\equiv 1\!\!\!\mod \frac{2v^2}{\overline{k}}
    \end{matrix}
  \end{align*}
  is equal to $2d(\overline{k})$.

  Let $8|k$.  Then the number of solutions $(v,z,x,y)\in\{1,\ldots,\overline{k}\}\times\{1,\ldots,\frac{v^2}{\overline{k}}\}\times\{1,3\}^2$ to the equations
  \begin{align*}
    \begin{matrix}
      \frac{v^2}{\overline{k}},\frac{\overline{k}}{v}\in\mZ,\,z\equiv\frac{k}{8}\!\!\!\mod2, & 4z^2\equiv 1\!\!\!\mod \frac{v^2}{2\overline{k}}
    \end{matrix}
  \end{align*}
  is equal to $2d(\overline{k})$.

  Let $4|k$.  Then the number of solutions $(v,z,x)\in\{1,\ldots,\frac{\overline{k}}{2}\}\times\{1,\ldots,\frac{2v^2}{\overline{k}}\}\times\{1,3\}$ to the equations
  \begin{align*}
    \begin{matrix}
      \frac{2v^2}{\overline{k}},\frac{\overline{k}}{2v}\in\mZ, & z^2\equiv 1\!\!\!\mod \frac{2v^2}{\overline{k}}
    \end{matrix}
  \end{align*}
  is equal to $d(\overline{k})$.

  Let $4|k$.  Then the number of solutions $(v,z,n)\in\{1,\ldots,\frac{\overline{k}}{2}\}\times\{1,\ldots,\frac{8v^2}{\overline{k}}\}\times\{0,1\}$ to the equations
  \begin{align*}
    \begin{matrix}
      \frac{2v^2}{\overline{k}},\frac{\overline{k}}{2v}\in\mZ, & z^2\equiv 1\!\!\!\mod \frac{4v^2}{\overline{k}}
    \end{matrix}
  \end{align*}
  is equal to $2d(\overline{k})$.
\end{lemma}
Again the details of the proof are to be found in~\cite{Gray2009}.\qed

\subsubsection{4 divides \texorpdfstring{$\overline{k}$}{kbar}}
As in the previous cases, every modular invariant with $4|\overline{k}$ is a simple current invariant.

Write $\overline{k}=2^mp$ where $p$ is odd and $m\geq 2$.  Then the subgroups of $\mZ_2\times\mZ_{\overline{k}}$ are given by 
\begin{align*}
  \begin{matrix}
    \mZ_2\times\mZ_l\cong\mZ_{2l},&l|p\\
    \mZ_2\times\mZ_{2l},&2l|\overline{k}\\
    \{0\}\times\mZ_{l}\cong\mZ_{l},&l|\overline{k}\\
    \langle(J,\frac{\overline{k}}{2l})\rangle\cong\mZ_{2l},&2l|\overline{k}.
  \end{matrix}
\end{align*}
Writing $\tau(G)$ for the number of degrees of freedom coming from discrete torsion of a subgroup $G$ of $\mZ_2\times\mZ_{\overline{k}}$ we find that the number of possible orbifolds of the diagonal partition function is
\begin{align*}
  N&=\sum_{G\leq \mZ_2\times\mZ_{\overline{k}}}\tau(G)=2\left(d(\overline{k})+d\left(\frac{\overline{k}}{2}\right)\right)
\end{align*}
The following lemma shows that this is precisely the number of simple current invariants when the level $k$ satisfies $4|\overline{k}$, proving that the Schellekens-Kreuzer orbifolds do indeed find all the modular invariants at these levels.
\begin{lemma}
  Let $4|\overline{k}$ and write $\overline{k}=2^{2r+\epsilon}p$ where $\epsilon\in\{0,1\},r>0$ and $p$ is odd.

  The number of solutions $(v,z,n)\in\{1,\ldots,\frac{\overline{k}}{2}\}\times\{1,\ldots,\frac{2v^2}{\overline{k}}\}\times\{0,1\}$ to the equations
  \begin{align*}
    \begin{matrix}
      \frac{v^2}{\overline{k}},\frac{\overline{k}}{2v}\in\mZ, & z^2\equiv 1\!\!\!\mod \frac{2v^2}{\overline{k}}
    \end{matrix}
  \end{align*}
  is equal to $2(4r-3+\epsilon)d(p)$.

  The number of solutions $(v,z,n)\in\{1,\ldots,\frac{\overline{k}}{2}\}\times\{1,\ldots,\frac{2v^2}{\overline{k}}\}\times\{0,1\}$ to the equations
  \begin{align*}
    \begin{matrix}
      \frac{2v^2}{\overline{k}}\in 2\mZ+1,\;\frac{\overline{k}}{2v}\in\mZ, & z^2\equiv 1\!\!\!\mod \frac{2v^2}{\overline{k}}
    \end{matrix}
  \end{align*}
  is equal to $2\epsilon d(p)$.

  The number of solutions $(v,z,n,m)\in\{1,\ldots,\overline{k}\}\times\{1,\ldots,\frac{2v^2}{\overline{k}}\}\times\{0,1\}^2$ to the equations
  \begin{align*}
    \begin{matrix}
      \frac{v^2}{\overline{k}}\in\mZ,\;\frac{\overline{k}}{v}\in 2\mZ+1, & z^2\equiv 1\!\!\!\mod \frac{4v^2}{\overline{k}}
    \end{matrix}
  \end{align*}
  is equal to $8d(p)$.\qed
\end{lemma}

\subsubsection{Simple Current Invariant Classification}
These counting results coupled with the explicit orbifolds given by Schellekens and Kreuzer~\cite{Kreuzer1994} can be summarised in the following theorem:
\begin{theorem}\label{countingtheorem}
  Every simple current $N=2$ unitary minimal partition function at level $k$ is realised via an orbifold (possibly with discrete torsion) of the diagonal partition function by a subgroup of the effective centre
  \begin{align*}
    \mathcal{C}\cong\begin{cases}
    \mZ_{2\overline{k}} & \text{if $k$ is odd,}\\
    \mZ_2\times\mZ_{2\overline{k}} & \text{if $4$ divides $k$,}\\
    \mZ_2\times\mZ_{\overline{k}} & \text{if $4$ divides $\overline{k}$.}
    \end{cases}
  \end{align*}
  The number of simple current invariants at each level $k\neq 2$ is given by\footnote{As discussed in section~\ref{k=2}, there are only five simple current invariants due to the identification $\mathcal{A}_2=\mathcal{D}_2$.}
  \begin{align}\label{N}
    N(k)=\begin{cases}
    2d(\overline{k}) & \text{if $k$ is odd,}\\
    5d(\overline{k}) & \text{if $4$ divides $k$,}\\
    2d(\overline{k})+2d\left(\frac{\overline{k}}{2}\right) & \text{if $4$ divides $\overline{k}$.}
    \end{cases}
  \end{align}
  where $d(n)$ is the number of divisors of $n$.\qed
\end{theorem}

\section{Conclusion}
We have reviewed Gannon's classification of the partition functions of the unitary $N=2$ minimal models and given the explicit results with a few minor errors corrected.  It is hoped that by making this list explicit, the less studied models therein may receive more attention.

The main result of this paper was to show that every one of these possible partition functions really does correspond to a full minimal SCFT, subject to assumption 1 and 2.  This is a large step towards completing the full classification of the unitary $N=2$ minimal models.

We also showed that Kreuzer and Schellekens' result that every simple current invariant is realised via an orbifolding of the diagonal partition function holds without the extra assumption of higher-genus modular invariant.  

This paper brings us tantalisingly close to the complete classification of the unitary $N=2$ minimal models.  To complete the classification, it must be shown that there is just one SCFT belonging to each partition function.

An alternative line of attack might be to approach the classification from the modular tensor category (see~\cite{Turaev1994,frs02}) point of view, or via the theory of nets of subfactors (see~\cite{Longo1994}).

It would also be satisfying to find some geometric classification of the minimal models in terms of singularities, analogous to the classification of the space-time supersymmetric  models in terms of simple singularities arising in their Landau-Ginzburg descriptions~\cite{Martinec1989,Vafa1989,Cecotti1993}.

\section*{Acknowledgement}
I would like to thank K. Wendland, T. Gannon and I. Runkel for useful discussions, and K. Wendland for posing the original problem.  Thanks also to the anonymous referees all gave very helpful and detailed comments.

I would like to acknowledge funding received from the University of Augsburg, the Deutsche Forschungsgemeinschaft (Grant Number WE 4340/1-1), the Engineering and Physical Sciences Research Council and the Heilbronn Institute at the University of Bristol.

\bibliographystyle{hplainbv}
\bibliography{/home/oliver/latex/bibtex/references}

\end{document}